\documentclass[article,twocolumn, floatfixnofootinbib]{revtex4}
\usepackage{filecontents}
\usepackage{hyperref}
\usepackage{graphicx} 
\usepackage{subfig}
\usepackage{color}

\newcommand{\ket}[1]{\left| #1 \right\rangle}
\newcommand{\bra}[1]{\left\langle #1 \right|}

\newcommand{\be}{\begin{equation}}
\newcommand{\ee}{\end{equation}}
\newcommand{\bea}{\begin{eqnarray}}
\newcommand{\eea}{\end{eqnarray}}

\usepackage{dcolumn}
\usepackage{bm}
\usepackage{amssymb,amsmath}
\usepackage{color}
\usepackage{float}
\usepackage{tikz}
\usetikzlibrary{arrows}

\definecolor{DarkGreen}{rgb}{0,0.6,0.2}

\newcommand*{\myeqref}[2][Eq.~]{%
  \hyperref[{#2}]{#1(\ref*{#2})}%
}
\def\equationautorefname#1#2\null{%
  Eq.#1(#2\null)%
}

\begin{document}
\title{
Two-photon entanglement in multi-qubit bi-directional waveguide QED}
\author{Imran M. Mirza}
\affiliation{Department of Physics, University of Michigan, Ann Arbor, Michigan 48109, USA}
\author{John C. Schotland}
\affiliation{Department of Mathematics and Department of Physics, University of Michigan, Ann Arbor, Michigan 48109, USA}

\begin{abstract}
We study entanglement generation and control in bi-directional waveguide QED driven by a two-photon Gaussian wavepacket. In particular, we focus on how increasing the number of qubits affects the overall average pairwise entanglement in the system. We also investigate how the presence of a second photon can introduce non-linearities, thereby manipulating the generated entanglement. In addition, we show that through the introduction of  chirality and small decay rates, entanglement can be stored and enhanced up to factors of $2$ and $3$, respectively. Finally, we analyze the influence of finite detunnings and time-delays on the generated entanglement. 
\end{abstract}

\maketitle

\section{Introduction}
Entanglement generation, maintenance and control lies at the heart of quantum teleportation, quantum communication, quantum cryptography and quantum computation \cite{nielsen2010quantum,aolita2015open}. Several quantum information processing protocols rely on controlled light-matter interactions which can entangle matter qubits through strongly or weakly interacting photons \cite{kok2010introduction}. In this context, cavity QED \cite{walther2006cavity} setups have been extensively studied with the aim of enabling entanglement transfer from photons to atoms \cite{ritter2012elementary,kastoryano2011dissipative,reiter2012driving}. However, for longer distance quantum communication, coupling of qubits with flying photonic mode reservoirs is a more advantageous approach. For this reason, the study of waveguide QED systems has garnered considerable recent attention~\cite{van2013photon,zheng2013waveguide}. In the standard setup of waveguide QED, qubits (atoms, quantum dots, nitrogen vacancy centers in diamond or superconducting Josephson junctions \cite{zoubi2014collective,arcari2014near,fu2008coupling,lalumiere2013input}) are placed near a waveguide (an optical fiber or a nanowire), and long-distance waveguide mediated qubit-qubit entanglement can be established. 

A related development is the study of two qubit entanglement in plasmonic waveguide systems~\cite{gonzalez2011entanglement,gangaraj2015transient,dzsotjan2010quantum}. Recently, Otten et al. has considered up to four plasmonically entangled quanum dots \cite{otten2015entanglement}. In such investigations, either an input coherent state pulse or a single photon generated within the system serves as a generator of entanglement. Interestingly, it has also been found that breaking the symmetry of qubit emission in chiral waveguides \cite{petersen2014chiral}) can lead to  enhancement of the generated entanglement~\cite{pichler2015quantum,gonzalez2015chiral}. 

The study of the propagation of quantum states of light through various material media is a subject of both fundamental and applied interest. A few examples reflecting this interest include: the observation of two-photon speckle patterns \cite{peeters2010observation}, radiative transport and scattering of two-photon entangled light \cite{schotland2016scattering,markel2014radiative}, two-photon imaging \cite{abouraddy2004entangled} and two-photon based quantum communications \cite{vaziri2002experimental}. Two-photon waveguide QED has also been investigated in recent years from the point of view of analyzing photon correlations and spectra. The problem of qubit-qubit entanglement generation has been relatively less studied, mainly due to the fact that a single photon can accomplish this task. However, the presence of a second photon in the waveguide can alter qubit-qubit entanglement in non-trivial ways. For instance, Ballestero et. al has shown that by launching two single-photon pulses from opposite ends of a waveguide, it is possible to manipulate the pattern of two-qubit entanglement by introducing a small time delay between the the pulses~\cite{gonzalez2014generation}. Moreover, such a scheme gives better control of the patterns of collapse and revival of qubit entanglement. 

Motivated by the above considerations, in this paper we study two-photon entanglement in multi-qubit waveguide QED systems.  In contrast to utilizing a weak laser pulse or other means to generate entanglement, here we consider a two-photon factorized Gaussian wavepacket pulse as an entanglement generator. Our main focus in this work is to examine how the presence of two simultaneously launched photons can introduce non-linearities in the qubits and thus affect the resulting multi-qubit entanglement. To this end, we derive and apply a two-photon bi-directional Fock state master equation. This approach differs from the most common techniques that are used to study the quantum dynamics of waveguide QED systems, namely Lehmberg type master equations \cite{lehmberg1970radiation,gonzalez2015chiral}, the real space formalism \cite{shen2009theory} and generalized input-out theory~\cite{caneva2015quantum}.

\begin{figure*}[t]
\includegraphics[width=6.85in,height=1.65in]{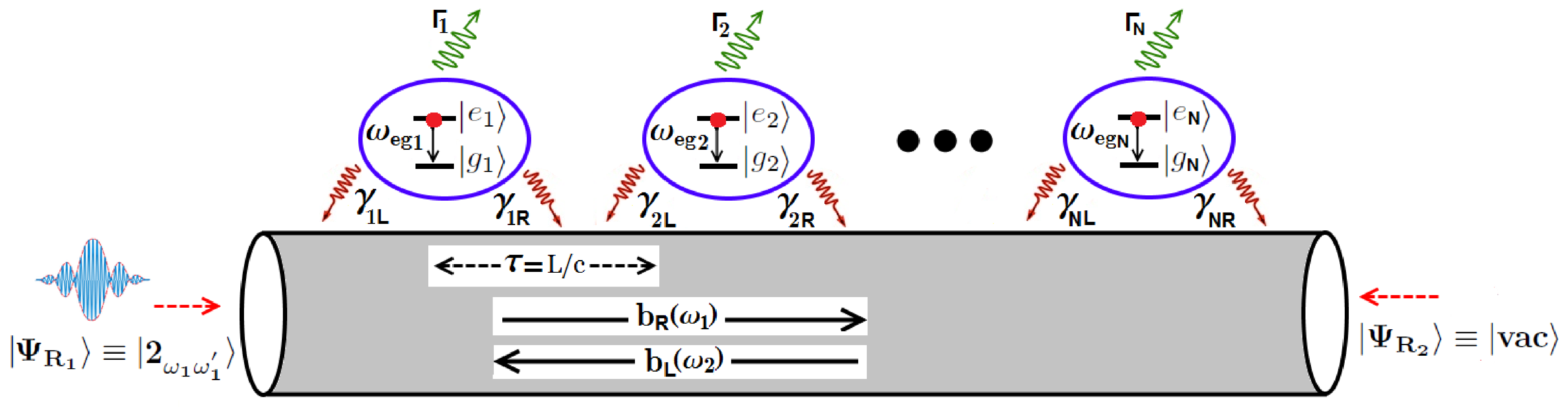}
\captionsetup{
  format=plain,
  margin=1em,
  justification=raggedright,
  singlelinecheck=false
}
\caption{A bi-directional waveguide QED setup. Atoms are separated by a distance $L$ which produces a time delay $\tau=L/c$ for the photon to propagate between any two consecutive atoms. The quantity $c\equiv v_{g}$ is the group velocity of the photons in the waveguide medium. Atoms can absorb incoming photons and then photons can either be emitted by the atoms into a free space channel (with rate $\Gamma_{i}$ for the $i$th atom) or in one of the two directions in the waveguide. Consequently, the coupling fraction parameter $\beta_{i}=(\gamma_{iL}+\gamma_{iR})/(\gamma_{iL}+\gamma_{iR}+\Gamma_{i})$ has been set equal to unity throughout this paper~\cite{gonzalez2015chiral}. Neglecting free space losses, the processes of photon emission and absorption result in the entanglement of the atoms in the chain. }\label{Fig1}
\end{figure*}

We find that for two qubit system, two photons produce a dip profile in the entanglement which diminishes as the number of qubits $N$ increases. In addition, the maximum value of entanglement shows a reduction of approximately 10\% for the $N=2$ case compared to the $N=5$ case. However, preferential directional emission of photons into the waveguide modes (chirality) can enhance the entanglement for the $N=5$ case by a factor of $2$. Similarly, the choice of smaller decay rates can improve the entanglement storage times by a factor $3$. Finally, we note that finite detuning between the peak frequency of the two-photon drive and the atomic transition frequency leads to a slight reduction in overall entanglement. Moreover, smaller delays support larger entanglement independent of $N$.

The remainder of this paper is organized as follows. In Sec.~II we describe the setup and dissipative dynamics of the system under study. Next, in Sec.~III we present and discuss our results. Finally, in Sec.~IV we formulate our conclusions. The derivation of the two-photon master equation we employ is presented in the Appendix.

\section{Theoretical description}
\subsection{Setup}
The system under investigation consists of a chain of two-level atoms (referred to as qubits) coupled to an optical waveguide, as shown in Fig.~1. The atomic transition frequency between the ground state $\ket{g_{i}}$ and excited state $\ket{e_{i}}$ of the $i$th atom is denoted by $\omega_{egi}$ and $\hat{\sigma}^{-}_{i}$ is the corresponding atomic lowering operator, for $i=1,2,\ldots,N$. The waveguide, which is assumed to be lossless and dispersionless, consists of two oppositely directed continua, referred to as left and right. Annihilation of a photon in right (left) going continuum is described by the operator $\hat{b}_{R}(\omega_{1})$ ($\hat{b}_{L}(\omega_{2})$). The nonvanishing commutation relations among these operators are of the form 
\begin{equation}
\begin{split}
&[\hat{b}_{R}(\omega_{1}),\hat{b}_{R}(\omega^{'}_{1})]=\delta(\omega_{1}-\omega^{'}_{1}) \ , \quad [\hat{\sigma}^{\dagger}_{i},\hat{\sigma}^{-}_{j}]=\hat{\sigma}_{zi}\delta_{ij} , \\
&[\hat{b}_{L}(\omega_{2}),\hat{b}_{L}(\omega^{'}_{2})]=\delta(\omega_{2}-\omega^{'}_{2})\ ,
\end{split}
\end{equation}
where $\hat{\sigma}_{zi}=\ket{e_{i}}\bra{e_{i}}-\ket{g_{i}}\bra{g_{i}}$. The system is taken to be driven from both ends of the waveguide. From the right hand side, it is driven by a reservoir $R_{2}$, which is initially in the pure vacuum state $\ket{\Psi_{R_{2}}}=\ket{vac}$. On the left hand side, the system is driven by an initial two-photon state $\ket{\Psi_{R_{1}}}$, which has the form
\begin{equation}
\label{eq:PsiR1}
\ket{\Psi_{R_{1}}}=\frac{1}{\sqrt{2}}\int^{\infty}_{0}\int^{\infty}_{0} g(\omega_{1},\omega^{'}_{1})\hat{b}^{\dagger}_{R}(\omega_{1})\hat{b}^{\dagger}_{R}(\omega^{'}_{1})\ket{vac}d\omega_{1}d\omega^{'}_{1} ,
\end{equation} 
where $g(\omega_{1},\omega^{'}_{1})$ is the spectral envelope of the two-photon wave packet. Note that normalization of $\ket{\Psi_{R_{1}}}$ requires that $\int^{\infty}_{0}\int^{\infty}_{0} |g(\omega_{1},\omega^{'}_{1})|^{2}d\omega_{1}d\omega^{'}_{1} = 1$.

\subsection{Dissipative dynamics and Master equation}

The above system is an open quantum system due to the interaction of the qubits with the waveguide continua. The dynamics of the state of the system is described by the following set of two-photon bi-directional Fock state master equations:
\begin{equation}
\label{eq:ME}
\begin{split}
&\hspace{-2mm}\frac{d\hat{\rho}_{s}(t)}{dt}=\mathcal{\hat{L}}[\hat{\rho}_{s}(t)]+
\sum^{N}_{i=1}\sqrt{2\gamma_{iR}}\Bigg(e^{ik_{0}d_{i}}g(t)[\hat{\rho}_{12}(t),\hat{\sigma}^{\dagger}_{i}]+h.c.\Bigg),\\
&\hspace{-3mm}\frac{d\hat{\rho}_{21}(t)}{dt}= \mathcal{\hat{L}}[\hat{\rho}_{21}(t)]+\sum^{N}_{i=1}\sqrt{2\gamma_{iR}}\Bigg(e^{ik_{0}d_{i}}g(t)[\hat{\rho}_{11}(t),\hat{\sigma}^{\dagger}_{i}]+h.c.\Bigg),\\
&\frac{d\hat{\rho}_{20}(t)}{dt}= \mathcal{\hat{L}}[\hat{\rho}_{20}(t)]+\sum^{N}_{i=1}\sqrt{2\gamma_{iR}}e^{ik_{0}d_{i}}g(t)[\hat{\rho}_{10}(t),\hat{\sigma}^{\dagger}_{i}],\\
&\frac{d\hat{\rho}_{11}(t)}{dt}= \mathcal{\hat{L}}[\hat{\rho}_{11}(t)]+\sum^{N}_{i=1}\sqrt{\gamma_{iR}}\Bigg(e^{ik_{0}d_{i}}g(t)[\hat{\rho}_{01}(t),\hat{\sigma}^{\dagger}_{i}]+h.c.\Bigg),\\
&\frac{d\hat{\rho}_{10}(t)}{dt}= \mathcal{\hat{L}}[\hat{\rho}_{10}(t)]+\sum^{N}_{i=1}\sqrt{\gamma_{iR}}e^{ik_{0}d_{i}}g(t)[\hat{\rho}_{00}(t),\hat{\sigma}^{\dagger}_{i}],\\
&\frac{d\hat{\rho}_{00}(t)}{dt}= \mathcal{\hat{L}}[\hat{\rho}_{00}(t)].
\end{split}
\end{equation}
h.c. stands for hermitian conjugate and the Liouvillian operator is defined by
\begin{equation}
\mathcal{\hat{L}}[\hat{\varrho}(t)]\equiv\mathcal{\hat{L}}_{cs}[\hat{\varrho}(t)]+\mathcal{\hat{L}}_{pd}[\hat{\varrho}(t)]+\mathcal{\hat{L}}_{cd}[\hat{\varrho}(t)]\nonumber,
\end{equation}
while
\begin{equation}
\begin{split}
&\mathcal{\hat{L}}_{cs}[\hat{\varrho}(t)]=-\frac{i}{\hbar}[\hat{H}_{sys},\hat{\varrho}(t)], \hat{H}_{sys}=\hbar\sum^{N}_{i=1}\Delta_{i}\hat{\sigma}^{\dagger}_{i}\hat{\sigma}^{-}_{i} \ , \\
&\hspace{-2mm}\mathcal{\hat{L}}_{pd}[\hat{\varrho}(t)]=-\sum^{N}_{i=1}\gamma_{iRL}(\hat{\sigma}^{\dagger}_{i}\hat{\sigma}^{-}_{i}\hat{\varrho}(t)-2\hat{\sigma}^{-}_{i}\hat{\varrho}(t)\hat{\sigma}^{\dagger}_{i}+\hat{\varrho}(t)\hat{\sigma}^{\dagger}_{i}\hat{\sigma}^{-}_{i}) \ , \\
&\mathcal{\hat{L}}_{cd}[\hat{\varrho}(t)]=-\sum^{N}_{i\neq j=1}(\sqrt{\gamma_{iR}\gamma_{jR}}\delta_{i>j}+\sqrt{\gamma_{iL}\gamma_{jL}}
\delta_{i<j})\\ \times
&\lbrace(\hat{\sigma}^{\dagger}_{i}\hat{\sigma}^{-}_{j}\hat{\varrho}(t)-\hat{\sigma}^{-}_{i}\hat{\varrho}(t)\hat{\sigma}^{\dagger}_{j})e^{-2\pi i D(i-j)}-h.c.\rbrace \ .
\end{split}
\end{equation}
Here $\delta_{i\lessgtr j}=1$ for all $i\lessgtr j$ and $\gamma_{iL} (\gamma_{iR})$ is the $i$th atom decay rate into the left (right) moving continuum. In addition, $d_{i}$ specifies the location of any $i$th atom, $\omega_{eg}$ is the common atomic transition frequency for all atoms and $D=L/\lambda_{0}$, with $\lambda_{0}=2\pi/k_{0}=2\pi v_{g}/\omega_{eg}$ the wavelength of the  emitted photon. The function $g(t)$ is a Gaussian obtained from the spectral profile function $g(\omega_{1},\omega_{2})$, as discussed in the next section. The derivation of the master equations~(\myeqref{eq:ME}) is presented in the Appendix.  

The first term on the right hand side of the master equation for $\hat{\rho}_{s}(t)$ describes the closed system dynamics, the second term (with the prefactor $(\frac{\gamma_{iR}+\gamma_{iL}}{2})$) represents the pure decay of energy from the atoms into the waveguide continua and finally the terms multiplied by $\sqrt{\gamma_{iR}\gamma_{jR}},\sqrt{\gamma_{iL}\gamma_{jL}}$ are the cooperative decay terms, with $j=1,2,\ldots ,N$. These cooperative decay terms originate from the coupling of the discrete energy levels of the atoms to the two common waveguide continua. The operators appearing in \myeqref{eq:ME} are of the form
\begin{subequations}
\begin{eqnarray}
\hspace{-5mm}\hat{\rho}_{21}(t)={\rm Tr_{R}}[\hat{U}(t,t_{0})\hat{\rho}_{s}(t)\ket{2_{\omega_{1}\omega^{'}_{1}}}\bra{\Psi^{1}}\hat{\rho}_{R_{2}}(t_{0})\hat{U}^{\dagger}(t,t_{0})],\hspace{2mm}\\
\hat{\rho}_{20}(t)={\rm Tr_{R}}[\hat{U}(t,t_{0})\hat{\rho}_{s}(t)\ket{2_{\omega_{1}\omega^{'}_{1}}}\bra{vac}\hat{\rho}_{R_{2}}(t_{0})\hat{U}^{\dagger}(t,t_{0})],\hspace{2mm}\\
\hspace{-3mm}\hat{\rho}_{11}(t)={\rm Tr}_{R}[\hat{U}(t,t_{0})\hat{\rho}_{s}(t)\ket{\Psi^{1}}\bra{\Psi^{1}}\hat{\rho}_{R_{2}}(t_{0})\hat{U}^{\dagger}(t,t_{0})],\hspace{2mm}\\
\hspace{-3mm}\hat{\rho}_{10}(t)={\rm Tr}_{R}[\hat{U}(t,t_{0})\hat{\rho}_{s}(t)\ket{\Psi^{1}}\bra{vac}\hat{\rho}_{R_{2}}(t_{0})\hat{U}^{\dagger}(t,t_{0})],\hspace{2mm}\\
\hspace{-3mm}\hat{\rho}_{00}(t)={\rm Tr}_{R}[\hat{U}(t,t_{0})\hat{\rho}_{s}(t)\ket{vac}\bra{vac}\hat{\rho}_{R_{2}}(t_{0})\hat{U}^{\dagger}(t,t_{0})],\hspace{2mm}
\end{eqnarray}
\end{subequations}
where $\hat{U}(t-t_{0})$ is the time evolution operator and $\ket{\Psi^{1}},\ket{2_{\omega_{1}\omega^{'}_{1}}}$ are the one- and two-photon reservoir states. Owing to their non-hermitian nature, the operators $\hat{\rho}_{21}(t), \ \hat{\rho}_{20}(t)$ and $\hat{\rho}_{10}(t)$ cannot be categorized as physical density operators, but they still obey the property $\hat{\rho}^{\dagger}_{21}(t)=\hat{\rho}_{12}(t), \ \hat{\rho}^{\dagger}_{20}(t)=\hat{\rho}_{02}(t)$ and $\hat{\rho}^{\dagger}_{10}(t)=\hat{\rho}_{01}(t)$. 

We remark that Combes et al. \cite{baragiola2012n} have derived a similar two-photon Fock state master equation using the machinery of quantum stochastic differential equations. However, we have not only followed a different route in derivation here, but our master equation also incorporates bi-directionalities, which is the central feature in waveguide QED problems. We note that the last three equations in \myeqref{eq:ME} can describe the complete evolution of the state of the system $\hat{\rho}_{s}(t)\equiv\hat{\rho}_{11}(t)$, if a single-photon wavepacket drives the system. Moreover, in the absence of any drive, the last master equation in \myeqref{eq:ME} is sufficient to describe the evolution of the atomic chain. Note that such a no-drive master equation can also be derived using second order perturbation theory under the application of the standard weak Born-Markov assumption, as originally described by Lehmberg~\cite{lehmberg1970radiation}. 

\section{Results and Discussion}

In this section we utilize the master equations \myeqref{eq:ME} to answer two questions. First, how do the atomic state populations evolve in response to the input drive? Second, how does the incoming two-photon wave packet generate and manipulate entanglement among the qubits? To set the stage, we begin with the simplest possible situation, namely a system consisting of only one atom.

\subsection{One atom system}

For this example, the system Hamiltonian becomes $\hat{H}_{sys}=\hbar\omega_{eg}\hat{\sigma}^{\dagger}\hat{\sigma}$ and we denote the decay rates by $\gamma_{1R}=\gamma_{1L}\equiv\gamma$.

\begin{figure}[h]
\includegraphics[width=2.95in,height=2.35in]{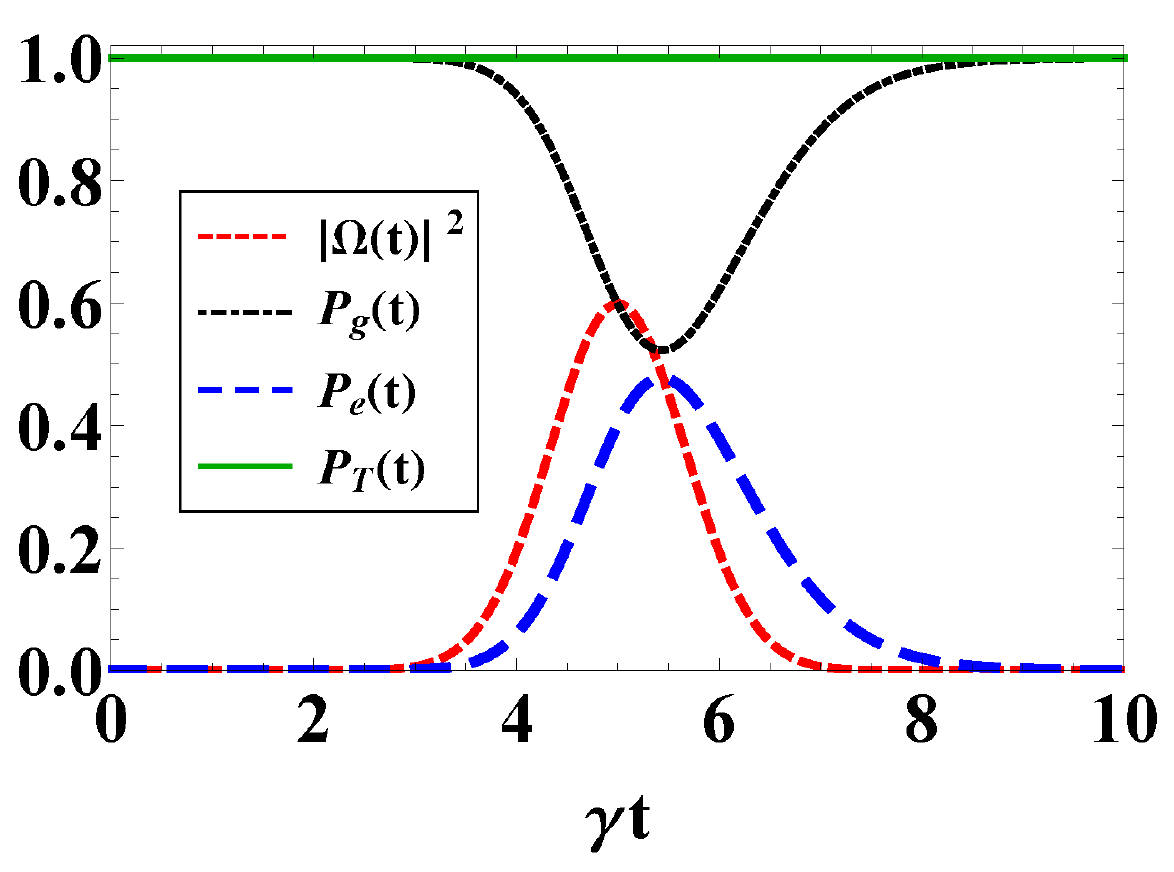}
\captionsetup{
  format=plain,
  margin=1em,
  justification=raggedright,
  singlelinecheck=false
}
 \caption{Time evolution of populations for a single side-coupled atom driven by a two-photon wave packet with time-dependent strength $\Omega(t)=\sqrt{2\gamma}g(t)$. Populations in the ground state ($P_{g}(t)$) and excited state ($P_{e}(t)$) are represented by black dotted dashed and blue longer dashed curves. The quantity $P_{T}(t)=P_g(t)+P_e(t)$ (green solid line) is shown to demonstrate conservation of the total population and the temporal pulse shape $|\Omega(t)|^{2}$ is shown in red with shorter dashing. The parameters used are $\overline{t}=5\gamma^{-1},\Delta t=1.5\gamma^{-1}$ with zero detuning between the peak frequency of the incoming wave packet and the atomic transition frequency.}\label{Fig2}
\end{figure}

We assume that initially the atom is in its ground state: $\hat{\rho}_{s}(t_{0})=\ket{g}\bra{g}$ and $\hat{\rho}_{21}(t_{0})=\hat{\rho}_{20}(t_{0})=\hat{\rho}_{10}(t_{0})=0$. As a useful consequence, we obtain $\hat{\rho}_{11}(t_{0})=\hat{\rho}_{00}(t_{0})=\ket{g}\bra{g}$. The spectral shape of the two-photon wave packet depends on the nature of the two-photon source. Here we assume that the two photons are produced by two independent single-photon sources such that the function $g(\omega_{1},\omega^{'}_{1})$ can be factorized in a symmetrized fashion using the Schmidt decomposition as 
\begin{equation}
g(\omega_{1},\omega^{'}_{1})=\frac{1}{2}\Bigg(g_{1}(\omega_{1})g_{2}(\omega^{'}_{1})+g_{2}(\omega_{1})g_{1}(\omega^{'}_{1})\Bigg).
\end{equation}
If we take each of the above factors to be Gaussian, then the two-photon wave packet will have a two Gaussian function product profile. In that case, the inverse Fourier transform of the spectral profile of any one of component functions is given by 
\begin{equation*}
g(t)=\frac{1}{\sqrt{2\pi}\Delta t}e^{-(t-\overline{t})^{2}/2(\Delta t)^{2}},
\end{equation*}
where $\overline{t}$ and $\Delta t$ specify the mean value and width of the Gaussian distribution, respectively. For experimental work related to the generation of two-photon states see references \cite{kumar2014controlling, halder2009nonclassical,garay2007photon}. 

In Fig.~2 we plot the atomic state populations under conditions when a two-photon wave packet strongly drives the atom ($|\Omega(t)|_{max}>\gamma$). The parameter choices have been made according to reference \cite{wang2011efficient} to obtain the highest probability of excitation. We note that as the wave packet enters the waveguide, after a small waiting time $\lesssim 0.5\gamma^{-1}$ the population $P_{e}(t)$ begins to grow. The highest value achieved by the excited state population is approximately 48\%. This value is smaller than the single atom excitation probability reported in Ref.~\cite{wang2011efficient}. The difference between the values can be attributed to the presence of bi-directional decays in our model. We also note that the overall temporal shape of the excited state population ($P_{e}(t)$) follows a symmetric behavior around its maximum value. Moreover, when the wavepacket amplitude $|\Omega(t)|$ vanishes at $t\sim 7\gamma^{-1}$, the atom still remains excited up to $40\%$ of its maximum value. The excited state population $P_{e}(t)$ takes a further time $t\sim\gamma^{-1}$ to completely diminish.

\subsection{Two atom chain and entanglement generation}

Next, we consider the case of two atoms. The presence of the second atom in the chain opens up the possibility of qubit-qubit entanglement. The two atoms in our system constitute a mixed state. The concurrence $\mathcal{C}(\hat{\rho}_{s})$ is an appropriate measure of entanglement in a bipartite mixed state \cite{wootters1998entanglement}. Following Wootters, we  define the concurrence $C(t)$ as
\begin{equation}\label{C}
\mathcal{C}(t)=\max\Bigg(0,\sqrt{\lambda_{1}}-\sqrt{\lambda_{2}}-\sqrt{\lambda_{3}}-\sqrt{\lambda_{4}}\Bigg),
\end{equation} 
where $\lambda_{i}$ are the eigenvalues (in descending order) of the spin flipped density matrix $\widetilde{\rho}_{s}=\hat{\rho}_{s}(\hat{\sigma}_{y}\otimes\hat{\sigma}_{y})\hat{\rho}^{\ast}_{s}(\hat{\sigma}_{y}\otimes\hat{\sigma}_{y})$, with $\hat{\sigma}_{y}$ being the Pauli spin flip operator. Note that $0\le\mathcal{C}\le 1$ and that $\mathcal{C}=1$ corresponds to a maximumally entangled state while $\mathcal{C}=0$ indicates a completely separable state.

In Fig.~3 we plot the population dynamics and the temporal profile of the entanglement. We see that the presence of the second atom means that there are now different possibilities available for the system to be excited. For instance, both atoms can be excited simultaneously ($P_{2}$) or only one of the atoms can be excited ($P_{1}$). Since both atoms are indistinguishable, we have plotted the sum of the probabilities of either one of the atoms to be excited. We observe that the maximum probability of either of the atoms to be excited is almost twice as high as the probability of both atoms to be excited simultaneously. Moreover, $P_{2}$ vanishes when the drive vanishes, while $P_{1}$ requires an additional time $t\sim\gamma^{-1}$ to vanish.

To facilitate our discussion of the concurrence, we first specify some notation and provide some details of our calculations. The relevant Hilbert space of the problem is spanned by the two qubit basis $\lbrace\ket{g_{1}g_{2}},\ket{e_{1}g_{2}},\ket{g_{1}e_{2}}, \ket{e_{1}e_{2}} \rbrace$, which we will refer to as $\lbrace\ket{1},\ket{2},\ket{3},\ket{4}\rbrace$. The density matrix consists of 16 elements. Through numerical integration of the equations of motion using the Runge-Kutta method of order 4 together with the initial condition $\hat{\rho}_{s}(t=0)=\ket{1}\bra{1}$, we find that all density matrix elements are real and 9 elements remain zero for all time. This leads us to the  simplified form of the spin flip density matrix:
\begin{equation}
\widetilde{\rho}_{s}(t) = 
 \begin{pmatrix}
  \rho^{2}_{4}(t)+\rho_{1}(t)\rho_{16}(t) & 0 & 0 & \rho_{1}\rho_{4} \\
  0 & 2\rho^{2}_{6}(t) & 2\rho^{2}_{6}(t) & 0 \\
   0 & 2\rho^{2}_{6}(t) & 2\rho^{2}_{6}(t) & 0  \\
 \rho_{1}\rho_{4} & 0 & 0 & \rho_{1}(t)\rho_{16}(t)
 \end{pmatrix},
\end{equation}
where $\rho_{1}(t)\equiv \bra{1}\hat{\rho}_{s}(t)\ket{1},\ \rho_{4}(t)\equiv \bra{1}\hat{\rho}_{s}(t)\ket{4}, \ \rho_{6}(t)\equiv \bra{2}\hat{\rho}_{s}(t)\ket{2},\ \rho_{16}(t)\equiv \bra{4}\hat{\rho}_{s}(t)\ket{4}$. Diagonalization of $\widetilde{\rho}_{s}(t)$ yields the following eigenvalues:
\begin{equation}
\begin{split}
&\lambda_{1}=0, \ \lambda_{2}=4\rho^{2}_{6}(t),\\
&\lambda_{3}=\rho_{1}(t)\rho_{16}(t)+\frac{1}{2}\rho_{4}(t)\Bigg(\rho_{4}(t)-\sqrt{\rho^{2}_{4}(t)+4\rho_{1}(t)\rho_{16}(t)}\Bigg),\\
&\lambda_{4}=\rho_{1}(t)\rho_{16}(t)+\frac{1}{2}\rho_{4}(t)\Bigg(\rho_{4}(t)+\sqrt{\rho^{2}_{4}(t)+4\rho_{1}(t)\rho_{16}(t)}\Bigg).
\end{split}
\end{equation}

\begin{figure*}
\centering
  \begin{tabular}{@{}cccc@{}}
    \includegraphics[width=3in, height=2.4in]{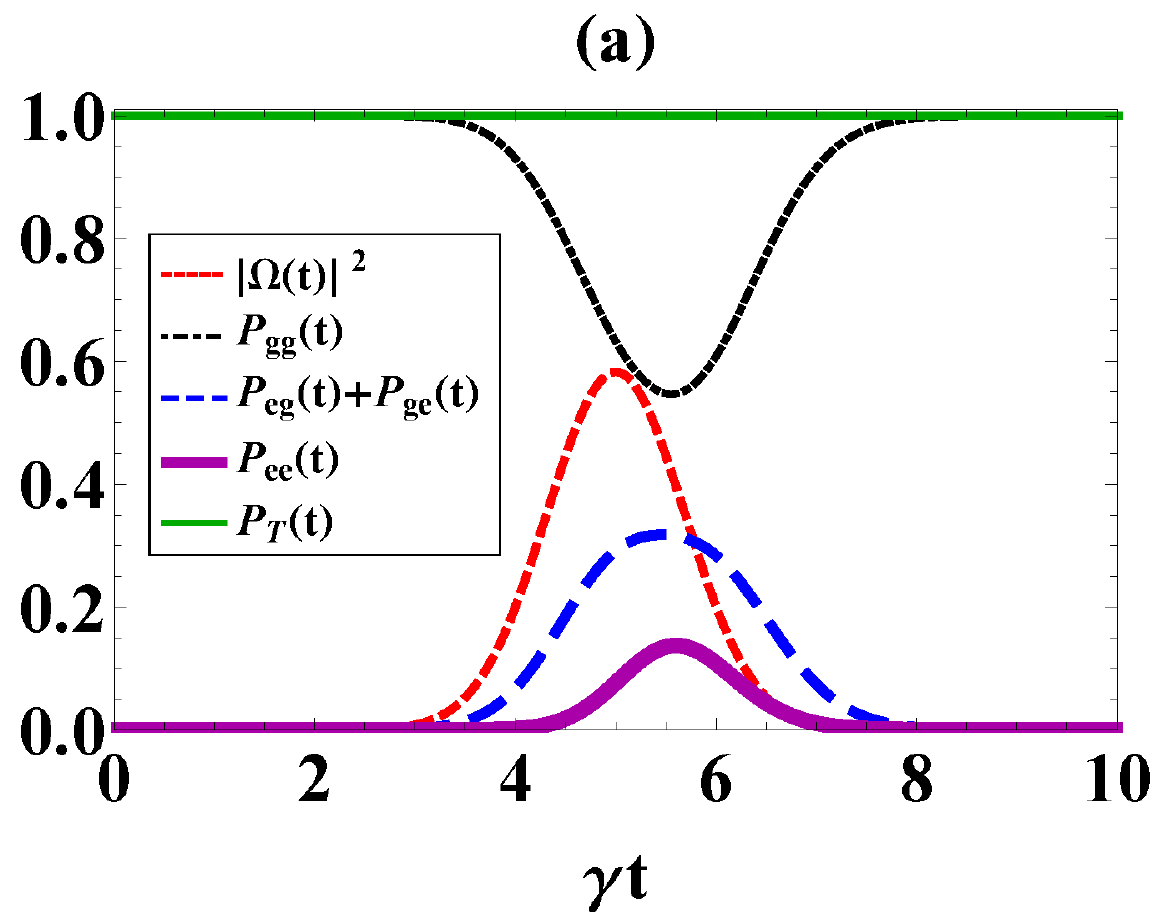}&
    \includegraphics[width=3in, height=2.4in]{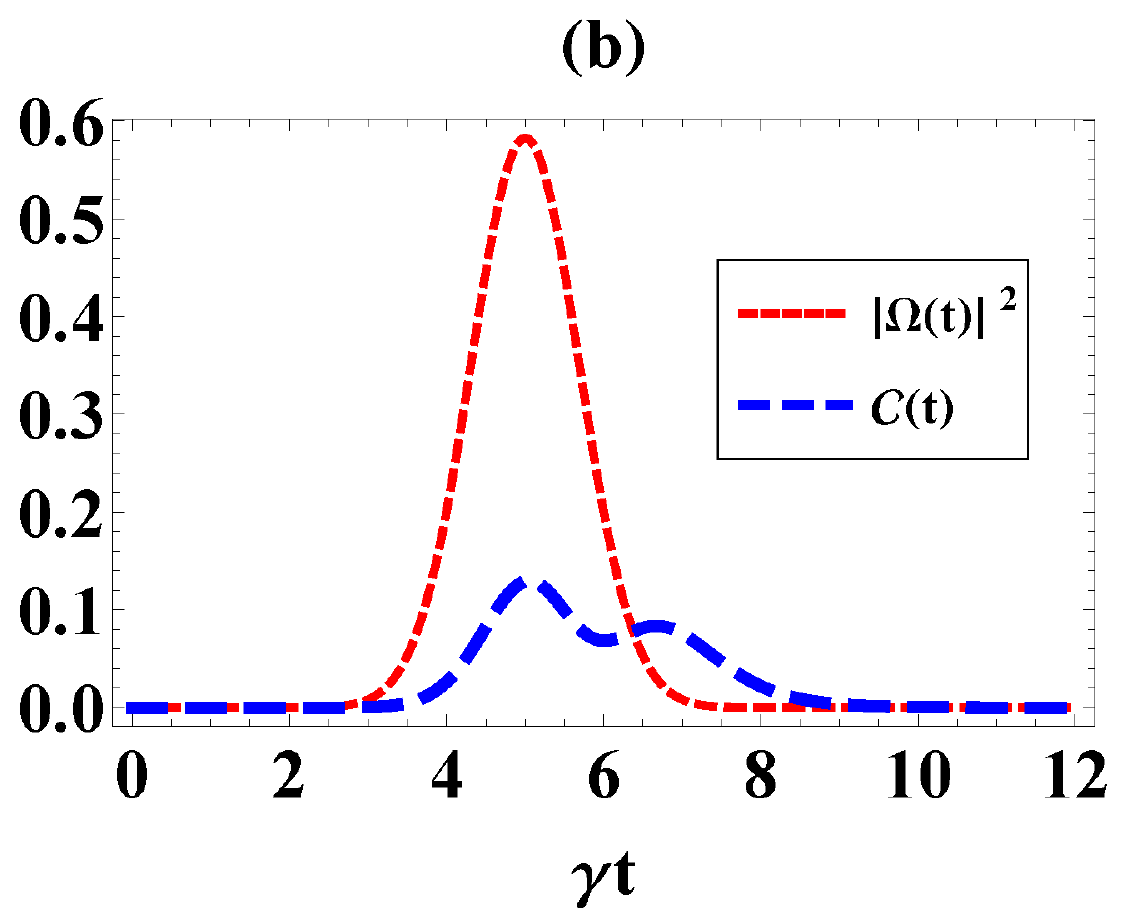}& 
  \end{tabular}
  \captionsetup{
  format=plain,
  margin=1em,
  justification=raggedright,
  singlelinecheck=false
}
\caption{Time evolution of (a) populations and (b) entanglement for a system of two identical atoms (qubits) coupled to a waveguide and driven by a two-photon wave packet. For simplicity, we have assumed all decay rates (pure and cooperative) to be equal. That is $\gamma_{1L}=\gamma_{2L}=\gamma_{1R}=\gamma_{2R}\equiv\gamma$. All other paramaters are the same as in Fig.~2. In the inset of Fig.~3(a) we use the notational convention that the first (second) slot specifies the state of the first (second) atom.}\label{Fig3}
\end{figure*}
\noindent
Inserting these eigenvalues into the definition of the concurrence, we obtain the required entanglement, which is plotted in Fig.~3(b). We find that the two photon wave packet generates entanglement between qubits while the highest value of the concurrence is 12\%. In addition, the temporal profile of entanglement shows a dip in between the two maxima. The first maximum appears at the time when the input drive reaches its highest value, at $t=5\gamma^{-1}$). The second maximum appears after a gap $t=2\gamma^{-1}$ when the wave packet has died out. We can explain these results by noting that when the two photon input drive enters the system, both atoms are excited simultaneously (we have neglected any time delays between the qubits). The atoms then gradually form a $(\ket{00}+\ket{11})/\sqrt{2}$ Bell state and the entanglement correspondingly increases. Later, one of the atoms loses a photon and the system forms a $(\ket{10}+\ket{01})/\sqrt{2}$ Bell state. The gap between the peaks in the concurrence can be interpreted as the time required for a single photon to be lost after shuttling between the qubits. Finally, at time $\sim t=9\gamma^{-1}$ the qubits becomes unentangled.

\subsection{Multi-qubit chain and average pairwise concurrence}

We now extend our study to include many-atoms in the chain. The main novelty of this section is the departure from a bipartite to a multipartite mixed state. We note that the entanglement quantification for multipartite mixed states is an open problem \cite{amico2008entanglement,horodecki2009quantum}. Here, we use the pairwise average concurrence as an entanglement measure \cite{amico2008entanglement,yonacc2007pairwise,wang2004entanglement,sarovar2010quantum}. To this end, we divide the system into all possible bipartite pairs of atoms, where the concurrence of the $i$th pair is given by $\mathcal{C}_{i}(t)$ and the total concurrence $\mathcal{C}(t)$ is given by: $\mathcal{C}(t)=(\sum^{n}_{i=1}\mathcal{C}_{i}(t))/n$ where $n=N/2$ is the total number of qubit pairs. We note that this definition of the concurrence has the same properties (including bounds on the highest and lowest values) as obeyed by the concurrence of a pair of atoms.

\begin{figure*}
\centering
  \begin{tabular}{@{}cccc@{}}
    \includegraphics[width=3in, height=2.4in]{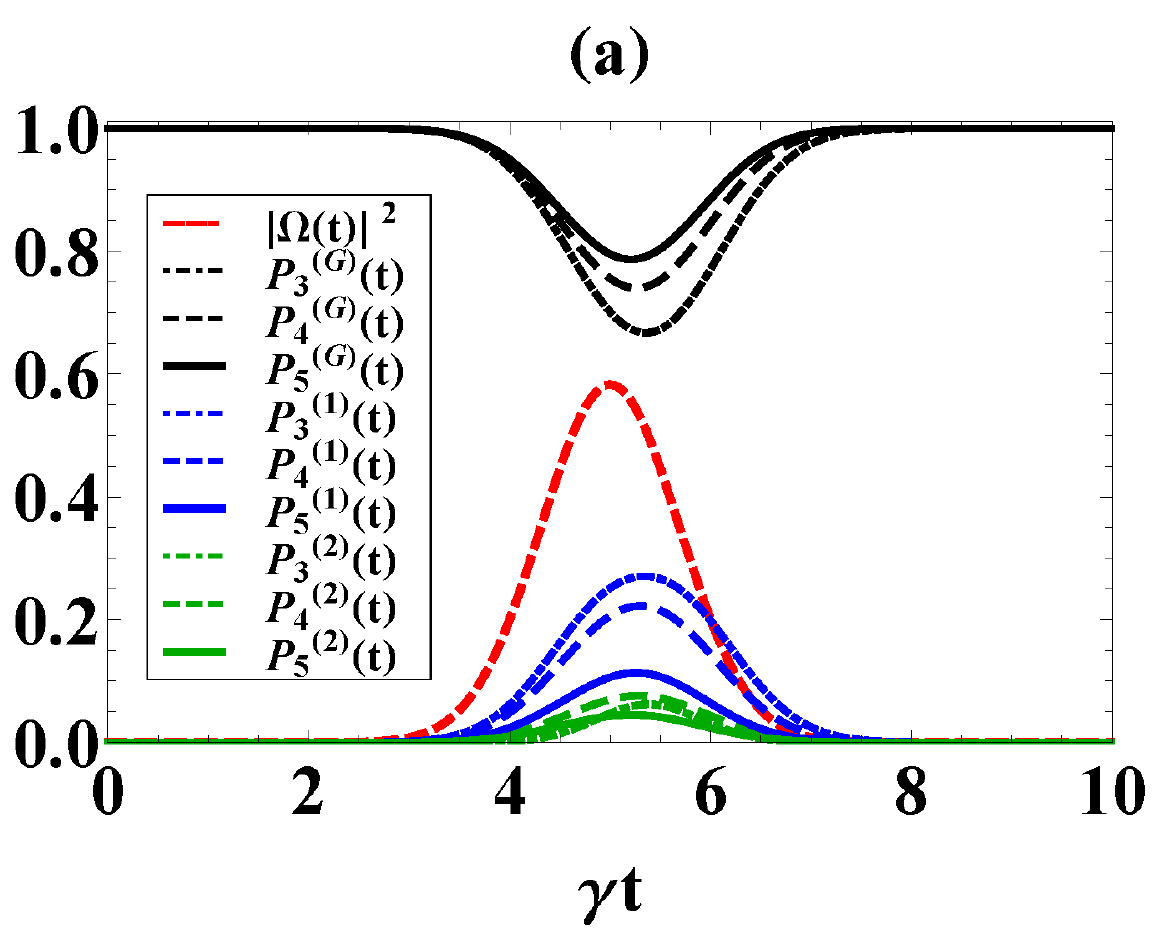}&
    \includegraphics[width=3in, height=2.4in]{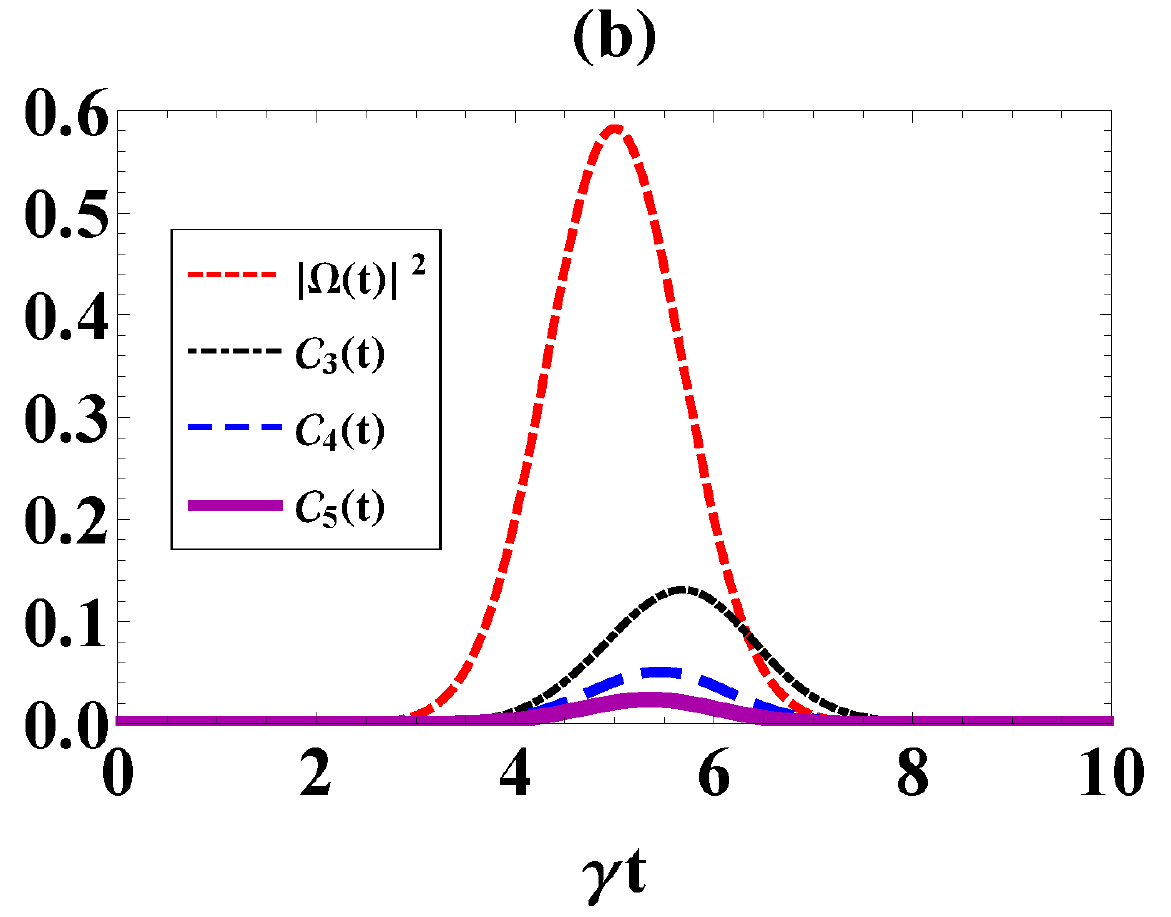}
  \end{tabular}
  \captionsetup{
  format=plain,
  margin=1em,
  justification=raggedright,
  singlelinecheck=false
}
  \caption{ Time evolution of (a) populations and (b) average pairwise concurrences for a system of 3, 4 and 5 qubits. All decay rates are chosen to be equal, with the remaining parameters the same as in Fig.~3. Here we use the notation that for $P^{(l)}_{k}(t)$ and $\mathcal{C}_{k}(t)$, $l=G,1,2$. In addition $G,1,2$ correspond to zero, one and two excitations in the system, while $k=3,4,5$ is the number of qubits in the chain.}\label{Fig4}
\end{figure*}

In Fig.~4(a) we present the population dynamics. We observe that as we increase the number of atoms in the chain, the probability that one or two  atoms are excited decreases. Moreover, the populations show a fast decay with increasing number of atoms. This observation can be attributed to the availability of more decay channels when the number of qubits in the system increases.

The pairwise entanglement (Fig.~4(b)) also attains smaller maxima and begins to decay quickly for an increasing number of qubits. Approximately $1/3$ and $1/2$ of the concurrence remains as we increase the number of qubits from three to four and four to five. In addition, the dip profile observed in the two qubit case also vanishes. 
This happens due to the availability of more qubits in the system which can absorb a photon emitted by one of the atoms. Thus later in time, it is possible to partially generate both type of Bell states ($(\ket{00}+\ket{11})/\sqrt{2}$ and $(\ket{10}+\ket{01})/\sqrt{2}$) in any one of the qubit pairs, which cannot happen in the two-qubit case.

\subsection{Small decays}

We now direct our attention to the case of small decay rates, which can be obtained by making use of reservoir engineering techniques (see for instance \cite{fedortchenko2014finite,schirmer2010stabilizing}). The main goal here is to optimize qubit decay rates so that the entanglement survival times can be increased. 
To this end we set the decay rate $\tilde{\gamma}=\gamma/10$. The corresponding results are presented in Fig.~5. In Fig.~5(a) we see that the single and double excitations remain in the system for more than double the time compared to Fig.~4. Similarly, we notice in Fig.~5(b) that the concurrence also survives longer.

\begin{figure*}
\centering
  \begin{tabular}{@{}cccc@{}}
   \includegraphics[width=2.3in, height=2in]{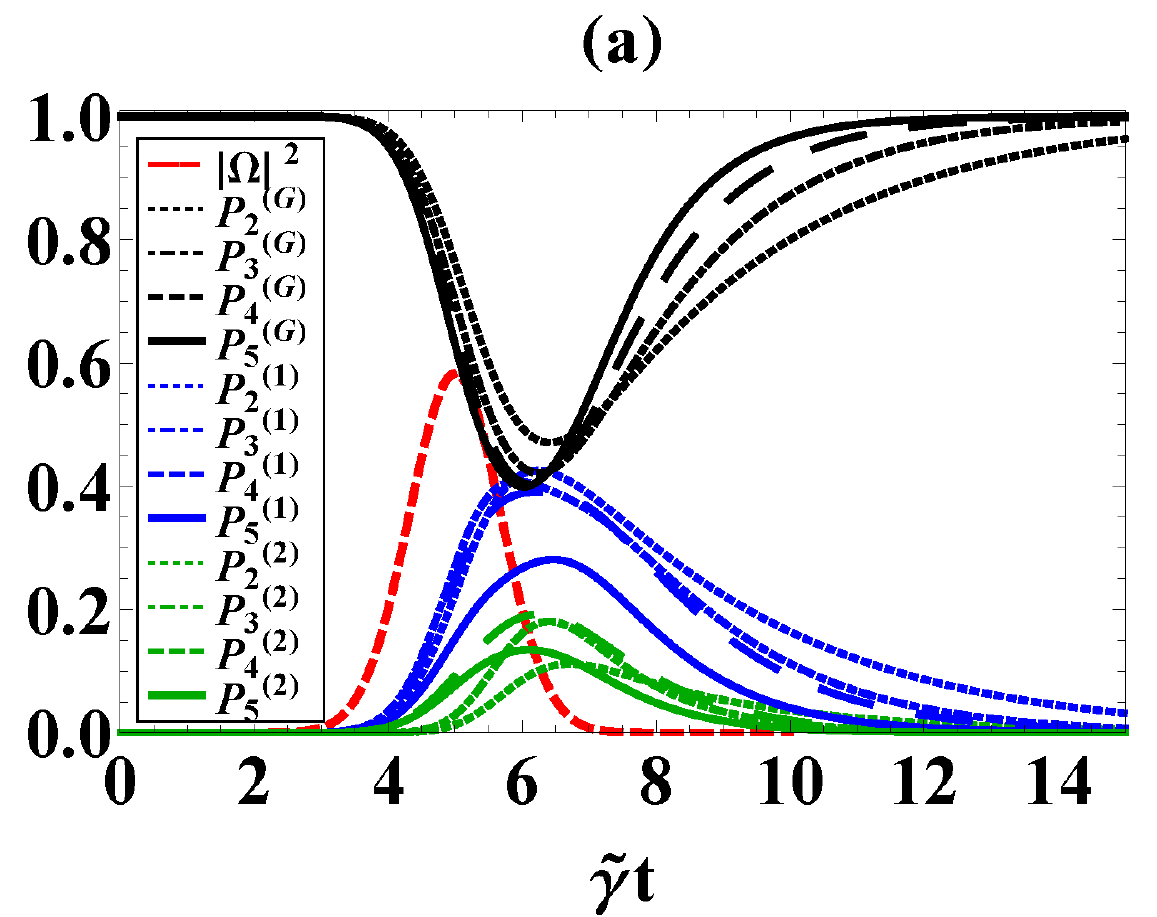}&
  \includegraphics[width=2.3in, height=2in]{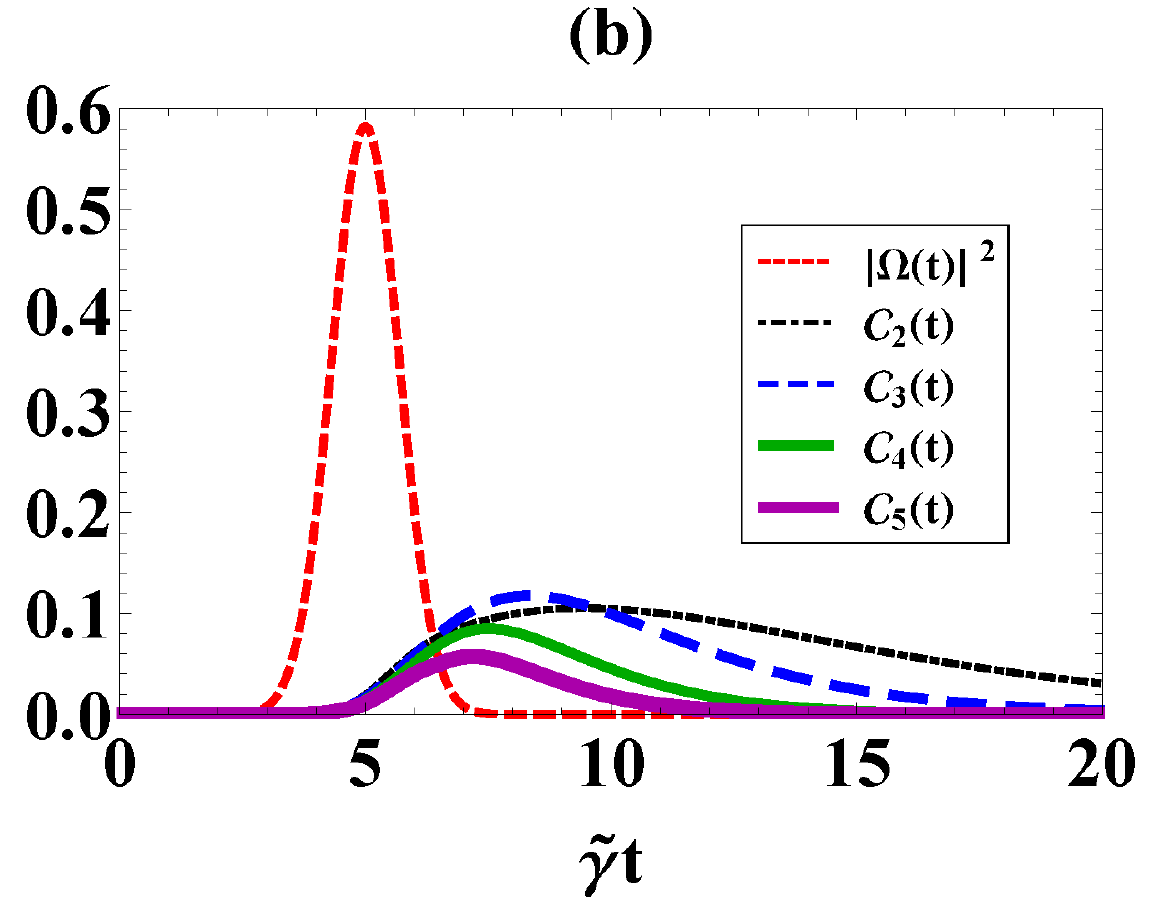}&
   \includegraphics[width=2.3in, height=1.96in]{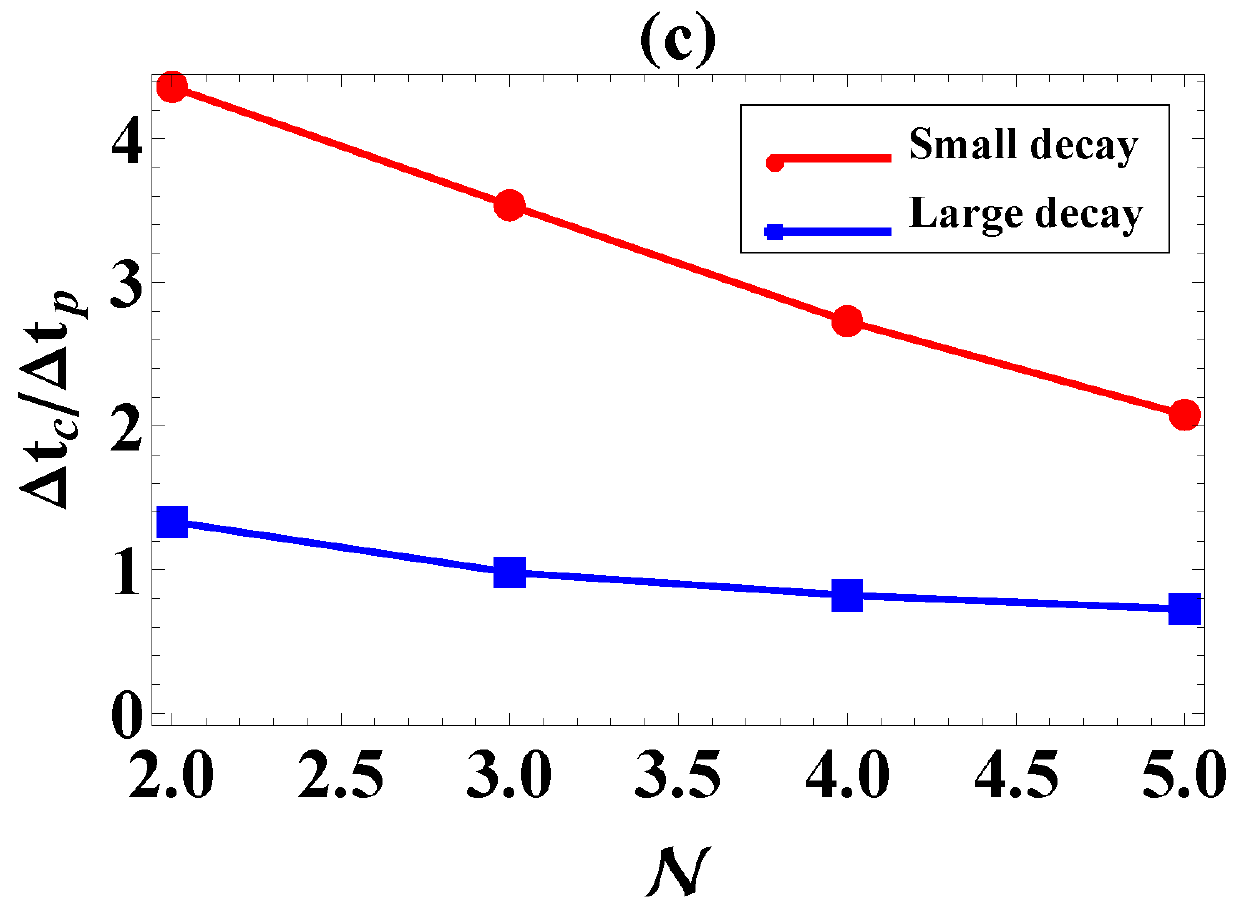}
  \end{tabular}
\captionsetup{
  format=plain,
  margin=1em,
  justification=raggedright,
  singlelinecheck=false
}
\caption{Influence of small decay rate on the time evolution of (a) populations and (b) pairwise concurrence for a 2, 3, 4 and 5 qubit system. All parameters are the same as used in Fig.~4 except we have chosen smaller coopertaive as well as pure decay rates i.e. $\tilde{\gamma}_{iL}=\tilde{\gamma}_{iR}=\tilde{\gamma}$ while $\tilde{\gamma}=0.1\gamma$. (c) Engantlement survival time $\Delta t_{c}$ as a function of pulse duration time $\Delta t_{p}$ plotted for both $\tilde{\gamma}$ and $\gamma$ scenarios.}\label{Fig5}
\end{figure*} 

The key point learned from Fig.~5 is that using small decay rates, the entanglement survival times can be increased without compromising the maximum entanglement achieved. This point is illustrated in Fig.~5(c), where the concurrence survival time $\Delta t_{c}$ is plotted as a function of pulse duration $\Delta t_{p}$ as the number of qubits in the chain is increased. We find that for small decay rates the entanglement survives for nearly twice as long compared to the results in Fig.~4. Finally, we point out that such a longer sustained entanglement is necessary in performing certain quantum information processing protocols (see Refs.\cite{clausen2011quantum, saglamyurek2015quantum,ding2015quantum} and applications mentioned therein).

\subsection{Chirality in cavity-waveguide coupling}

There have been exciting recent developments in the subject of preferential atomic emission in waveguide QED systems due to spin-orbit interaction of light (chirality)~\cite{petersen2014chiral,ramos2014quantum,le2014anisotropy,pichler2015quantum}. In this section we analyze the ways in which chirality can impact the entanglement. To this end, we set the parameters $\gamma_{R}=5\gamma_{L}$, $\gamma_{iR}=\gamma_{R}$ and $\gamma_{iL}=\gamma_{L}$ for all $i$. Note that this choice of parameters lies within the recently acheieved 90\% directionalities and 98\% atom-waveguide coupling strengths in photonic crystal systems~\cite{sollner2015deterministic}.

\begin{figure*}
\centering
  \begin{tabular}{@{}cccc@{}}
   \includegraphics[width=2.3in, height=2in]{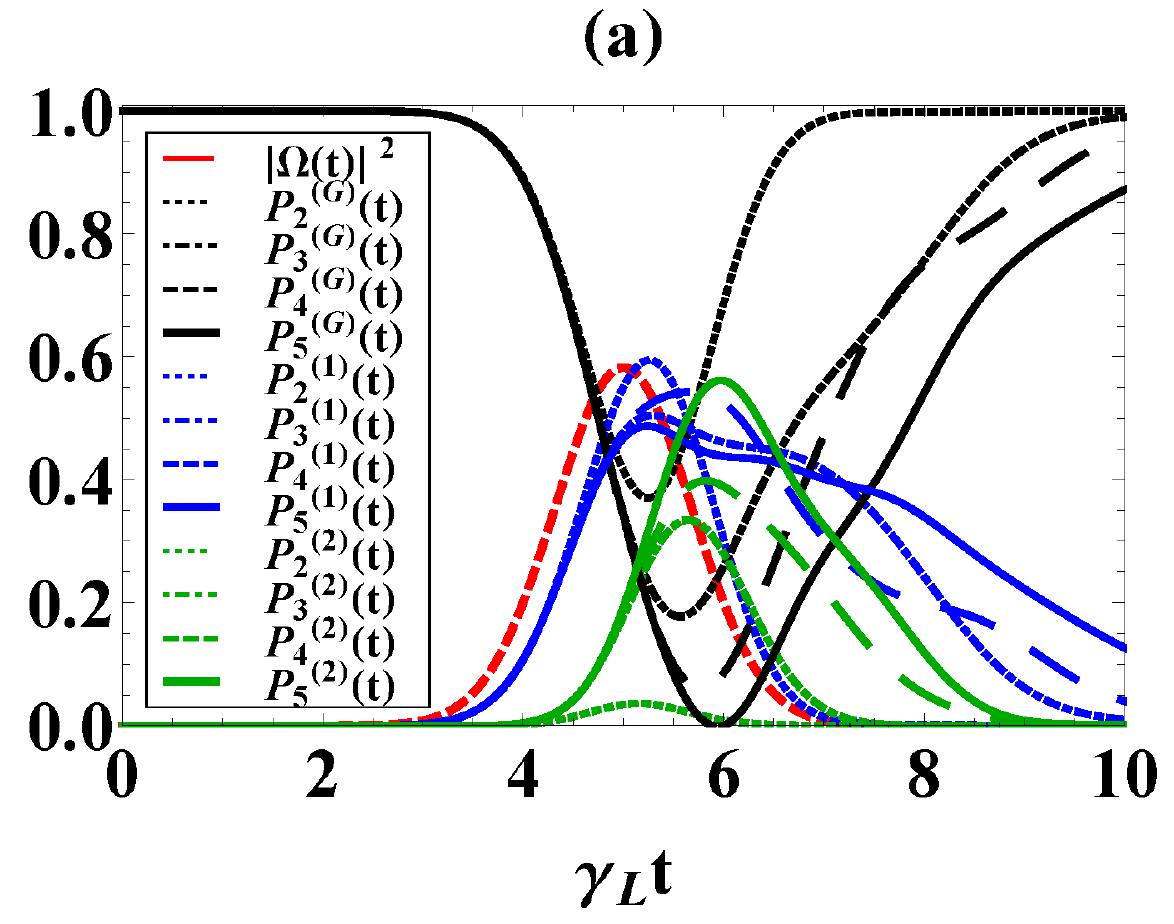}&
  \includegraphics[width=2.3in, height=2in]{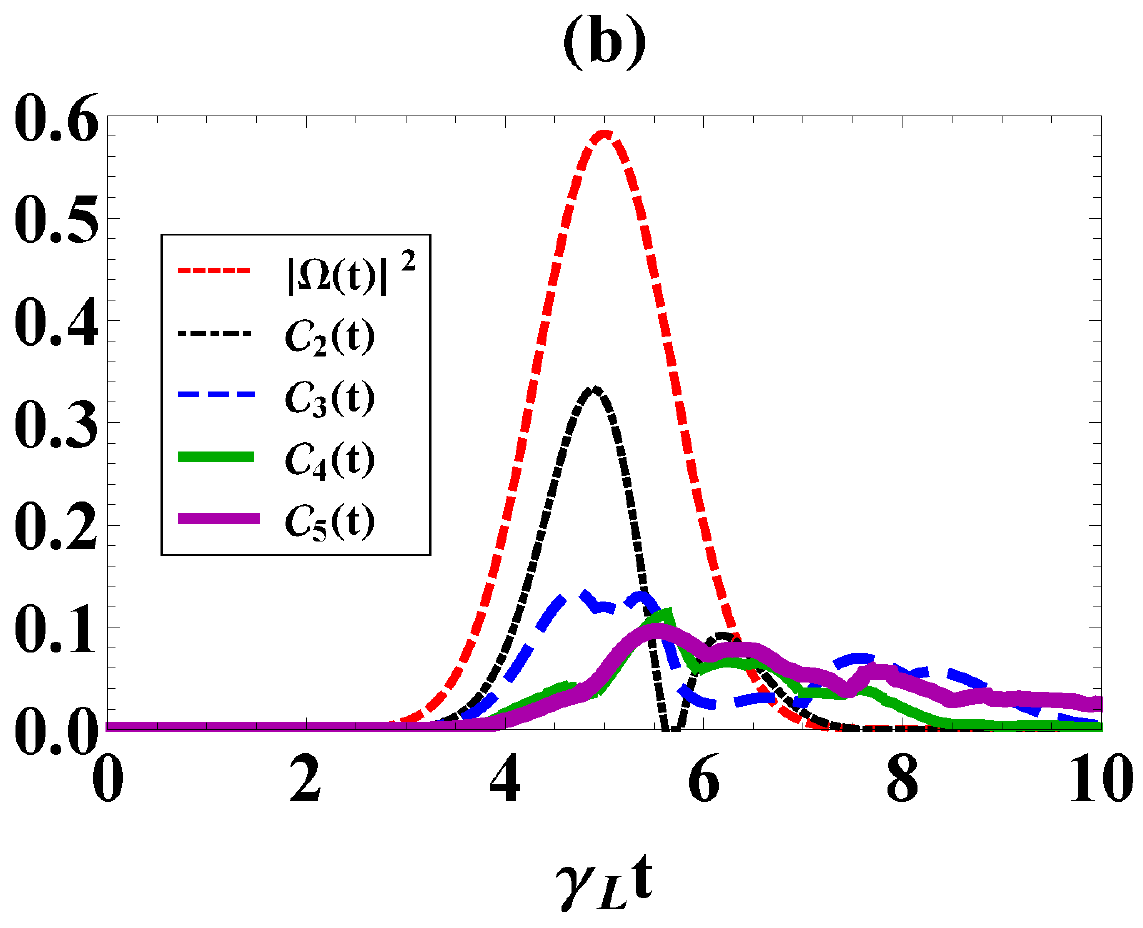}&
   \includegraphics[width=2.3in, height=1.9in]{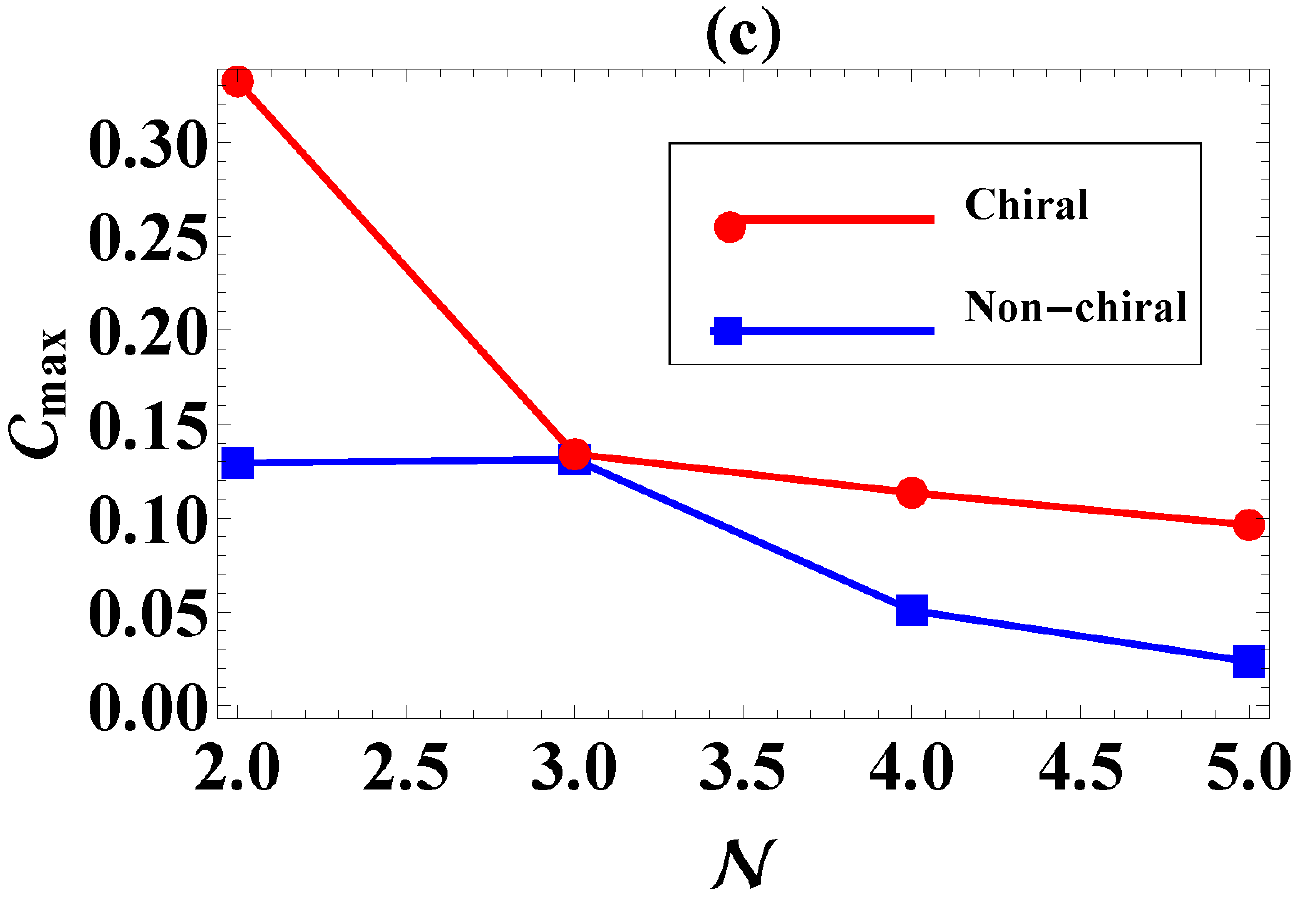}
  \end{tabular}
\captionsetup{
  format=plain,
  margin=1em,
  justification=raggedright,
  singlelinecheck=false
}
 \caption{Illustrating the effect of breaking the symmetry in the atomic emission directions for a multi-qubit waveguide system. (a) Population and (b) entanglement dynamics. We have chosen the parameters $\gamma_{1L}=\gamma_{2L}=\gamma_{3L}=\gamma_{4L}=\gamma_{5L}$ $\equiv\gamma_{L}$ (similarly for all $\gamma_{iR}$, for $i=2,3,4,5$) except $\gamma_{iR}/\gamma_{iL}=5$. The remainder of the parameters are the same as in Fig.~2. In order to emphasize the fact that chirality enhances the maximum entanglement generated in the system, we have also plotted the maximum concurrence ($\mathcal{C}_{max}$) as a function of $N$ for both chiral and no-chiral ($\gamma_{iR}=\gamma_{iL}=1$) situations. }\label{Fig5}
\end{figure*}

As shown in Fig.~6, there is a marked effect of chirality on the populations as well as on the entanglement dynamics of the system. In Fig.~6(a), we see that the single excitation populations become twice as large as in the non-chiral case (compare to Fig.~4(a)) and there is a corresponding increase in the survival time. Most interestingly, the two photon excitation population becomes almost five times larger than in the non-chiral case, especially when there are higher numbers of qubits in the chain. Finally, we note that in the populations plot for the 5 qubit chain, at the time $t\sim 6\gamma^{-1}_{L}$ the system is fully excited and the ground state population vanishes. This novel feature is a pure chirality effect.
 
The above described enhancement in the populations also translates into higher and longer survival of the entanglement, as shown in Fig.~6(b). We note that independent of the number of qubits, the pairwise concurrence displays an irregular oscillatory behavior. Moreover for the case of two qubits, the phenomenon of entanglement death and revival \cite{mazzola2009sudden,xu2010experimental} appears. Along with the longer storage of entanglement, which can also be obtained using small decay rates, the main advantage chirality offers is the enhancement of the achievable maximum entanglement. This point is emphasized in Fig.~6(c) where we have plotted the maximum entanglement for the chiral and non-chiral cases. We see that for all $N$, under chiral conditions, the maximum concurrence provides an upper bound on the non-chiral maximum concurrence, and for some $N$ can cause the entanglement to be even twice as large as in the non-chiral case. Note that Ballestero et al \cite{gonzalez2015chiral} have reported that chirality can enhance the single-photon entanglement in a two-qubit waveguide system by a factor of approximately 1.5. We, on the other hand, we have shown that using two-photon Gaussian wavepackets leads to a twice enhancement in entanglement in two-qubit chiral waveguide systems. 

\subsection{Detuning and Delays}

We now consider the situation in which $\omega_{p}$ (two photon wavepacket peak frequency) is slightly detuned from $\omega_{eg}$. In particular, we focus on how detunning alters the on-resonance entanglement among qubits. In Fig.~7(a) we plot our results. We notice that in all cases, detuning preserves the qualitative features of the concurrence but the entanglement is slightly reduced. Beginning with the two-atom case, we observe that detuning reduces the maximum entanglement by a factor of $\sim 8$\%, while the dip profile is preserved. Moreover, the difference between $\mathcal{C}_{2N}$ and $\mathcal{C}_{2D}$ tends to be greater for the second maximum, which causes the concurrence to die out quickly. As we increase the number of qubits in the chain, we note that the maximum entanglement difference becomes $\sim 17$\%, $10$\% and $6$\% for the 3, 4 and 5 qubit cases, respectively.

\begin{figure*}[t]
\centering
  \begin{tabular}{@{}cccc@{}}
    \includegraphics[width=3in, height=2.4in]{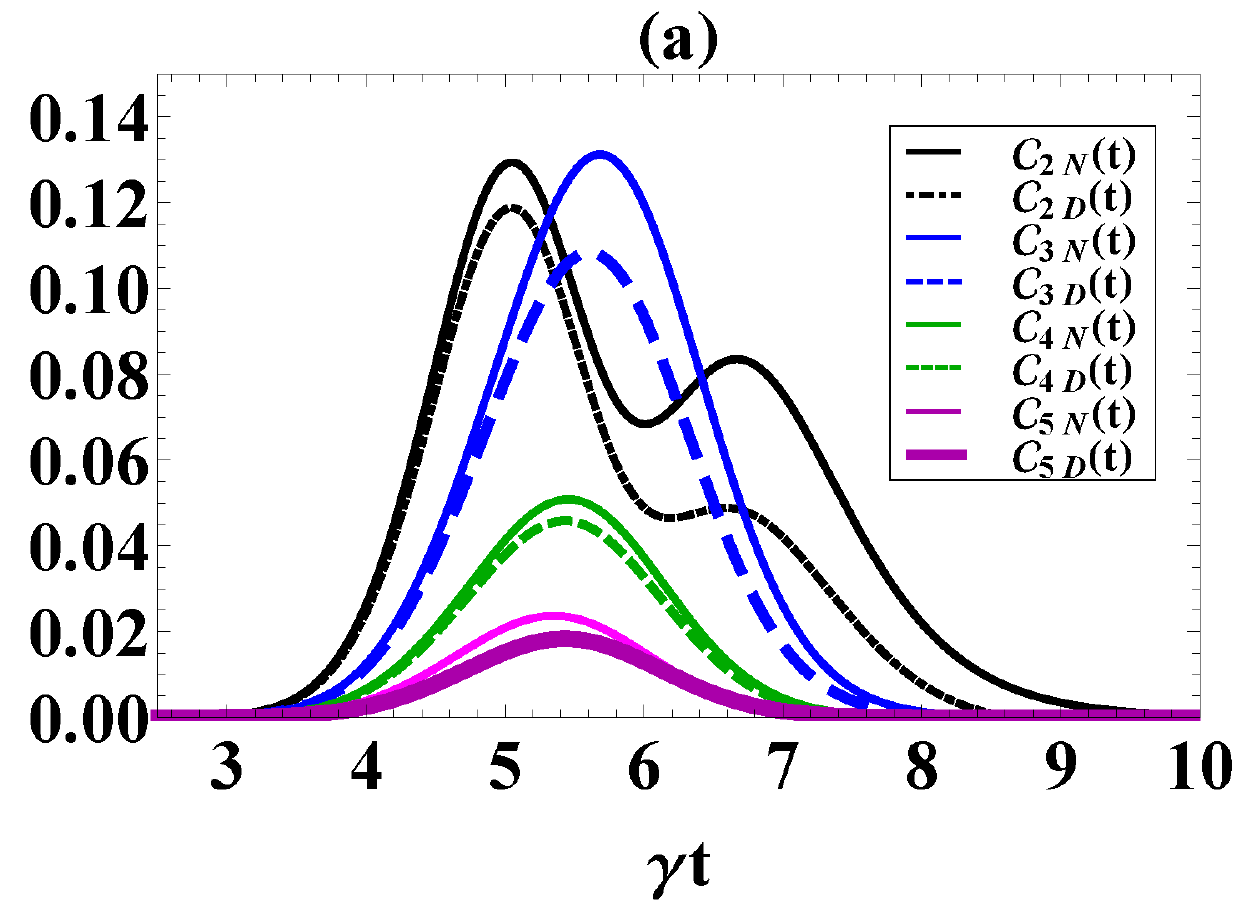}&
    \includegraphics[width=3in, height=2.4in]{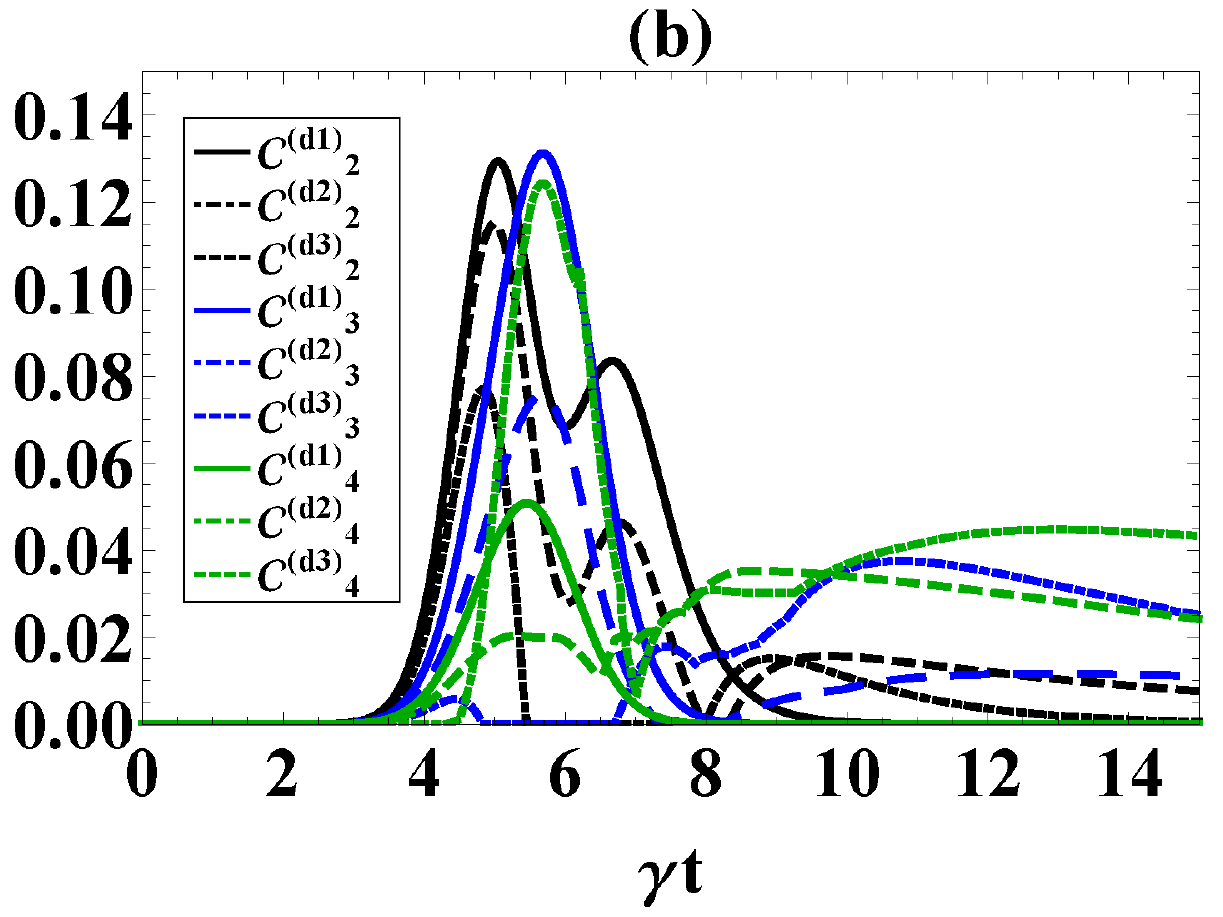}
  \end{tabular}
  \captionsetup{
  format=plain,
  margin=1em,
  justification=raggedright,
  singlelinecheck=false
}
 \caption{(a) Effect of detuning on the entanglement evolution. All atoms in the chain are assumed to have the same resonant frequency $\omega_{eg}$, which is $0.5\gamma$ detuned from $\omega_{p}$. We are using the notation that $\mathcal{C}_{kN}$ and $\mathcal{C}_{kD}$ are the concurrence for the no-detuning and finite detuning cases, respectively, where $k=2,3,4,5.$ (b) Entanglement dynamics in the presence of time delays between the atoms. Three cases are plotted, namely $d_{1}=L, d_{2}=L/8$ and $d_{3}=L/16$. The remaining parameters are the same as in Fig.~2.}
\label{Fig7}
\end{figure*}

Next, we consider the effect of delays on entanglement. Although we have neglected the time delays between the qubits originating form the input-output relations (see the Appendix ), there are still phases that appear in the atom-waveguide interaction Hamiltonian which carry information about the atomic positions. The two-photon master equation we have derived retains memory of the reservoir state and hence has a non-Markovian structure (see Appendix). To this end, we have considered three cases of inter-atomic separations, keeping in mind the already reported condition ($\gamma D \leq v_{g}$) for Markovian dynamics to hold~\cite{fang2015waveguide,tufarelli2014non}. In Fig.~7(b) we study entanglement in the  presence of finite delays. In the two-qubit case, we observe that as the separation is reduced from $L/8$ to $L/16$ the oscillatory profile survives, but the dip is suppressed. Note that even for $L/16$, the dip vanishes completely and a dark period of entanglement between $t=5.5$ to $8\gamma^{-1}$ emerges. Around $t=8\gamma^{-1}$, the entanglement revives and after quickly reaching a maximum value it decays steadily. 

In the case of $N\geq2$ qubits, the smallest separation produces an overall larger entanglement accompanied by dark and bright periods of entanglement. For instance, for the $N=4$ example $\mathcal{C}_{max}\sim 0.125$, which is more than two times greater than the maximum entanglement in the largest separation case ($\sim 0.055$). Note that in all of these plots, the entanglement decay and revival patterns originate from the delays. Therefore, through proper tuning of qubit-waveguide interaction phases, one can control the entanglement revival times which may find applications in quantum networks based on multi-qubit waveguide QED.

\section{Conclusions}
In summary, we have calculated and analyzed two-photon induced entanglement in multi-qubit waveguide QED. Using a bi-directional Fock state master equation together with the average pairwise concurrence as a measure of entanglement, we found that an incoming two-photon wave packet can entangle 2 qubits up to $\approx 12$\% and that the entanglement survives even after the passage of the driving wavepacket. However, the maximum pairwise entanglement decreases and decays rapidly as the number of qubits increases. The entanglement survival times can be increased by a factor of two with almost the same maximum entanglement, by using smaller decay rates $\tilde{\gamma}=\gamma/10$. 

The maximum value of the entanglement decreases by increasing the number of qubits. This problem can be mitigated by making use of chiral waveguide networks. We concluded that by choosing a five times larger decay rate in the direction of the incoming two-photon wave packet, we can achieve up to a factor of two greater maximum entanglement compared to the non-chiral situation. 

Finally, we studied the effects of detunings and delays. We found that detuning does not change the overall temporal profile of the entanglement, but a slight reduction in entanglement does occur. In contrast, delays independent of the value of $N$, produce death and revival patterns of entanglement, where the smallest inter-qubit separations support an overall higher entanglement. 

\section*{Acknowledgments}
This work was supported by the NSF Grants DMR-1120923, DMS-1115574 and DMS-1108969.

\setcounter{equation}{0}
\makeatletter 
\section*{Appendix: Two photon master equation}
\renewcommand{\theequation}{A\arabic{equation}}
In this Appendix we derive the two-photon master equation that we use throughout the paper. We begin by dividing a system of $N$ qubits coupled to a bi-directional waveguide into $N$ subsystems, where each subsystem consists of a single atom (or the $i$th quantum system with arbitrary system operators $\hat{c}_{i}$ and $\hat{X}_{i}$. In case of atoms $\hat{c}_{i}$ and $\hat{X}_{i}$ will represent $\hat{\sigma}^{-}_{i}$).

We begin by deriving the dissipative dynamics of the first subsystem in the Heisenberg picture. The Hamiltonian of the first system interacting with two reservoirs is given by
\begin{widetext}
\begin{equation}
\begin{split}
&\hat{H}=\hat{H}_{sys1}+\int^{\infty}_{-\infty} \hbar\omega_{1}\hat{b}^{\dagger}_{R}(\omega_{1})\hat{b}_{R}(\omega_{1})d\omega_{1}+\int^{\infty}_{-\infty} \hbar\omega_{2}\hat{b}^{\dagger}_{L}(\omega_{2})\hat{b}_{L}(\omega_{2})d\omega_{2}-i\hbar\sqrt{\frac{\gamma_{1R}}{2\pi}}\int^{\infty}_{-\infty} (e^{ik_{0}d_{1}}\hat{c}^{\dagger}_{1}\hat{b}_{R}(\omega_{1})\\
&-e^{-ik_{0}d_{1}}\hat{b}^{\dagger}_{R}(\omega_{1})\hat{c}_{1})d\omega_{1}-i\hbar\sqrt{\frac{\gamma_{1L}}{2\pi}}\int^{\infty}_{-\infty} (e^{-ik_{0}d_{1}}\hat{c}^{\dagger}_{1}\hat{b}_{L}(\omega_{2})-e^{ik_{0}d_{1}}\hat{b}^{\dagger}_{L}(\omega_{2})\hat{c}_{1})d\omega_{2}.
\end{split}
\end{equation}
\end{widetext}
Next, we transform to the Heisenberg picture, where the right-moving continuum evolves as
\begin{equation}
\frac{d\hat{b}_{R}(\omega_{1};t)}{dt}=-i\omega_{1}\hat{b}_{R}(\omega_{1};t)+\sqrt{\frac{\gamma_{1R}}{2\pi}}e^{-ik_{0}d_{1}}\hat{c}_{1}(t).
\end{equation}
For some initial time $t_{0}$, we obtain the solution at time $t$ in the form
\begin{equation}
\begin{split}
&\hat{b}_{R}(\omega_{1};t)=\\
&\hat{b}_{R}(\omega_{1};t_{0})e^{-i\omega_{1}(t-t_{0})}+\sqrt{\frac{\gamma_{1R}}{2\pi}}e^{-ik_{0}d_{1}}\int^{t}_{t_{0}}c_{1}(t^{'})e^{-i\omega_{1}(t-t^{'})}dt^{'}.
\end{split}
\end{equation}
Similarly, for the left-moving continuum we find
\begin{equation}
\begin{split}
&\hat{b}_{L}(\omega_{2};t)=\\
&\hat{b}_{L}(\omega_{2};t_{0})e^{-i\omega_{2}(t-t_{0})}+\sqrt{\frac{\gamma_{1L}}{2\pi}}e^{ik_{0}d_{1}}\int^{t}_{t_{0}}c_{1}(t^{'})e^{-i\omega_{2}(t-t^{'})}dt^{'}.
\end{split}
\end{equation}
Next, we introduce a system operator $\hat{X}_{1}(t)$ which obeys the Heisenberg equations of motion
\begin{equation}
\begin{split}
&\frac{d\hat{X}_{1}(t)}{dt}=\\
&\frac{-i}{\hbar}[\hat{X}_{1}(t),\hat{H}_{sys1}]-\sqrt{\frac{\gamma_{1R}}{2\pi}}(e^{ik_{0}d_{1}}\int^{\infty}_{-\infty}[\hat{X}_{1}(t),c^{\dagger}_{1}(t)]\hat{b}_{R}(\omega_{1})d\omega_{1}\\
&+h.c.)
-\sqrt{\frac{\gamma_{1L}}{2\pi}}(e^{-ik_{0}d_{1}}\int^{\infty}_{-\infty}[\hat{X}_{1}(t),c^{\dagger}_{1}(t)]\hat{b}_{L}(\omega_{2})d\omega_{2}+h.c.).
\end{split}
\end{equation}
After eliminating the continua in the above equation, we arrive at
\begin{widetext}
\begin{equation}
\begin{split}
&\frac{d\hat{X}_{1}(t)}{dt}=\frac{-i}{\hbar}[\hat{X}_{1}(t),\hat{H}_{sys1}]-[\hat{X}_{1}(t,\hat{c}^{\dagger}_{1}(t)]\Bigg(\sqrt{\gamma_{1R}}e^{ik_{0}d_{1}}\hat{b}^{(1R)}_{in}(t)+\sqrt{\gamma_{1L}}e^{-ik_{0}d_{1}}\hat{b}^{(1L)}_{in}(t)+(\frac{\gamma_{1R}+\gamma_{1L}}{2})\hat{c}_{1}  \Bigg)\\
&+\Bigg(\sqrt{\gamma_{1R}}e^{-ik_{0}d_{1}}\hat{b}^{\dagger(1R)}_{in}(t)
+\sqrt{\gamma_{1L}}e^{ik_{0}d_{1}}\hat{b}^{\dagger(1L)}_{in}(t)+(\frac{\gamma_{1R}+\gamma_{1L}}{2})\hat{c}^{\dagger}_{1}  \Bigg)[\hat{X}_{1}(t),\hat{c}_{1}(t)].
\end{split}
\end{equation}
\end{widetext}
The above quantum Langevin equation~\cite{gardiner2004quantum}) describes the dissipative dynamics of the first subsystem in the Heisenberg picture. In writing this equation we have identified two input operators
\begin{subequations}
\begin{eqnarray}
\hat{b}_{in}^{(1R)}(t)=\frac{1}{\sqrt{2\pi}}\int^{\infty}_{-\infty}\hat{b}_{R}(\omega_{1},t_{0})e^{-i\omega_{1}(t-t_{0})}d\omega_{1} , \\
\hat{b}_{in}^{(1L)}(t)=\frac{1}{\sqrt{2\pi}}\int^{\infty}_{-\infty}\hat{b}_{L}(\omega_{2},t_{0})e^{-i\omega_{2}(t-t_{0})}d\omega_{2}.
\end{eqnarray}
\end{subequations}
The input operators obey the causality condition  manifested by the commutation relation: $[\hat{b}^{(1j)}_{in}(t),\hat{b}^{\dagger(1j)}_{in}(t^{'})]=\delta(t-t^{'})$, with $j=R,L$. We note that corresponding to each input operator there exists an output operator with corresponding input-output relations. For system~1 coupled to the right- and left-moving continua, the input-output relation takes the form
\begin{subequations}
\begin{eqnarray}
\hat{b}^{(1R)}_{out}(t)=\hat{b}^{(1R)}_{in}(t)+\sqrt{\gamma_{1R}}e^{-ik_{0}d_{1}}\hat{c}_{1}(t),\\
\hat{b}^{(1L)}_{out}(t)=\hat{b}^{(1L)}_{in}(t)+\sqrt{\gamma_{1L}}e^{ik_{0}d_{1}}\hat{c}_{1}(t),
\end{eqnarray}
\end{subequations}
where $t_{1}$ is some future time and we have defined the output operator as
\begin{equation}
\hat{b}_{out}^{(1R/L)}(t)=\frac{1}{\sqrt{2\pi}}\int^{\infty}_{-\infty}\hat{b}_{R/L}(\omega_{1},t_{1})e^{-i\omega_{1}(t-t_{1})}d\omega_{1}.
\end{equation} 
Following along the same lines, one can derive a quantum Langevin equation obeyed by each member in the atomic chain.
\begin{figure*}[t]
\includegraphics[width=7.1in,height=2.1in]{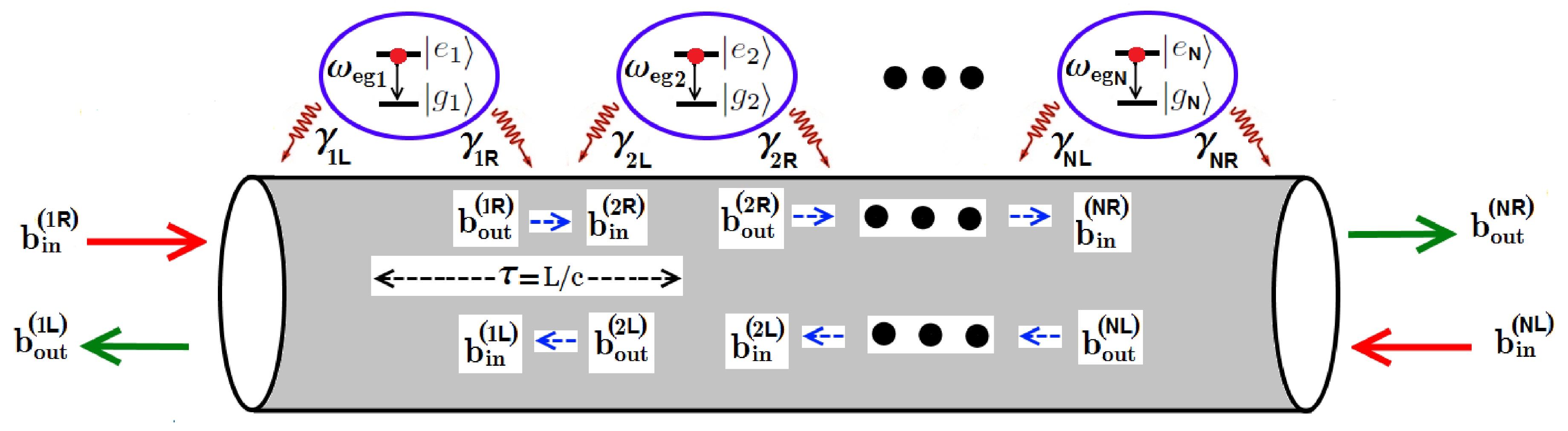}
\captionsetup{
  format=plain,
  margin=1em,
  justification=raggedright,
  singlelinecheck=false
}
 \caption{Bi-directional coupling among qubits caused by intra-waveguide input-output relations, where the output from one atom serves as the input to another.}\label{FigS1}
\end{figure*}
  Next, we note that the output from one subsystem feeds into the nearest subsystems as a time-delayed input. For instance, for the case of two subsystems, we have
\begin{subequations}
\begin{eqnarray}
\hat{b}_{in}^{(2R)}(t)=\hat{b}_{out}^{(1R)}(t-\tau)=\hat{b}_{in}^{(1R)}(t-\tau)+\sqrt{\gamma_{1R}}e^{-ik_{0}d_{1}}\hat{c}_{1}(t-\tau) \ , \nonumber\\
\hat{b}_{in}^{(1L)}(t)=\hat{b}_{out}^{(2L)}(t-\tau)=\hat{b}_{in}^{(2L)}(t-\tau)+\sqrt{\gamma_{1R}}e^{ik_{0}d_{1}}\hat{c}_{2}(t-\tau)\nonumber \ .
\end{eqnarray}
\end{subequations}
Thus, for $N$ subsystems we arrive at the following form of the combined Langevin equations:
\begin{widetext}
\begin{equation}
\begin{split}
&\frac{d\hat{X}(t)}{dt}=\frac{-i}{\hbar}[\hat{X}(t),\hat{H}_{sys}]
-\sum^{N}_{i=1}\Bigg\lbrace[\hat{X}(t),\hat{c}^{\dagger}_{i}(t)]\Bigg(\sqrt{\gamma_{iR}}e^{ik_{0}d_{i}}\hat{b}^{(iR)}_{in}(t)+\sqrt{\gamma_{iL}}e^{-ik_{0}d_{i}}\hat{b}^{(iL)}_{in}(t)+(\frac{\gamma_{iR}+\gamma_{iL}}{2})\hat{c}_{i}\\
& +\sum^{N}_{j\neq i=1}e^{ik_{0}(d_{i}-d_{j})}(\sqrt{\gamma_{iR}\gamma_{jR}}\delta_{i>j}\hat{c}_{j}(t)
+\sqrt{\gamma_{iL}\gamma_{jL}}\delta_{i<j}\hat{c}_{j}(t))\Bigg)+h.c.\Bigg\rbrace.
\end{split}
\end{equation}
\end{widetext}
Here we have neglected all intra-atom time delays under the assumption that the system evolves on a time scale much slower than the time a photon takes to travel between the atoms. That is $\omega_{egi},\gamma_{il}\ll {1}/{\tau}={L}/{c}, \ l=R,L$.
Next, we transform to the Schr\"odinger picture using the identity 
\begin{equation}
{\rm Tr}_{S\oplus R}\Bigg[\frac{d\hat{X}(t)}{dt}\hat{\rho}(t_{0})\Bigg]={\rm Tr}_{S}\Bigg[\hat{X}(t_{0})\frac{d\hat{\rho}_{s}(t)}{dt}\Bigg],
\end{equation}
where $\hat{\rho}_{s}(t)$ is the system reduced density matrix we are seeking. Using the cyclic property of the trace, we finally arrive at the master equation
\begin{widetext}
\begin{equation}
\begin{split}
&\frac{d\hat{\rho}_{s}(t)}{dt}=-\frac{i}{\hbar}\Bigg[\hat{H}_{sys},\hat{\rho}_{s}(t)\Bigg]-\sum^{N}_{i=1}(\frac{\gamma_{iR}+\gamma_{iL}}{2})(\hat{c}^{\dagger}_{i}\hat{c}_{i}\hat{\rho}_{s}(t)-2\hat{c}_{i}\hat{\rho}_{s}(t)\hat{c}^{\dagger}_{i}+\hat{\rho}_{s}(t)\hat{c}^{\dagger}_{i}\hat{c}_{i})-\sum^{N}_{i\neq j=1}(\sqrt{\gamma_{iR}\gamma_{jR}}\delta_{i>j}+\sqrt{\gamma_{iL}\gamma_{jL}}
\delta_{i<j})\times\\
&(e^{-ik_{0}(d_{i}-d_{j})}(\hat{c}^{\dagger}_{i}\hat{c}_{j}\hat{\rho}_{s}(t)-\hat{c}_{i}\hat{\rho}_{s}(t)\hat{c}^{\dagger}_{j})-h.c.)-{\rm Tr}_{S\oplus R}\Bigg[\sum^{N}_{i=1}\Bigg(\sqrt{\gamma_{iR}}(e^{ik_{0}d_{i}}[\hat{X}(t),\hat{c}^{\dagger}_{i}(t)]\hat{b}^{(1R)}_{in}(t)\hat{\rho}(t_{0})
-e^{-ik_{0}d_{i}}\hat{b}^{\dagger(1R)(t)}_{in}\times\\
&[\hat{X}(t),\hat{c}_{i}(t)]\hat{\rho}(t_{0}))-
\sqrt{\gamma_{iL}}(e^{-ik_{0}d_{i}}[\hat{X}(t),\hat{c}^{\dagger}_{i}(t)]\hat{b}^{(NL)}_{in}(t)\hat{\rho}(t_{0})-e^{ik_{0}d_{i}}\hat{b}^{\dagger(NL)(t)}_{in}[\hat{X}(t),\hat{c}_{i}(t)]\hat{\rho}(t_{0}))\Bigg)\Bigg].
\end{split}
\end{equation}
\end{widetext}

We now focus our attention to the input terms. First, we notice a considerable simplification, namely that the left moving continuum is initialy in a vacuum state. As a result, all terms involving the $\hat{b}^{(NL)}_{in}(t)$ operator vanish:
\begin{equation}
\begin{split}
&{\rm Tr}_{S\oplus R}\Bigg[[\hat{X}(t),\hat{c}^{\dagger}_{i}(t)]\hat{b}^{(NL)}_{in}(t)\hat{\rho}(t_{0})\Bigg]=\\
&{\rm Tr}_{S\oplus R}\Bigg[[\hat{X}(t),\hat{c}^{\dagger}_{i}(t)]\hat{\rho}_{s}(t_{0})\otimes\hat{\rho}_{R1}(t_{0})\otimes\hat{b}^{(NL)}_{in}(t)\ket{vac}\bra{vac}\Bigg]\\
&=0,
\end{split}
\end{equation}
where we have taken the initial system-environment state to be factorizable. Next, we focus on the right-moving continuum input terms; these do not vanish due to the presence of two-photons in the initial state of this reservoir:
\begin{equation}
\hat{b}^{(1R)}_{in}(t)\ket{2_{\omega_{1}\omega^{'}_{1}}}=2\int^{\infty}_{0} g_{R}(\omega_{1},t)\hat{b}^{\dagger}_{1}(\omega_{1})\ket{vac}d\omega_{1},
\end{equation}
where 
\begin{equation}
g_{R}(\omega_{1},t)=\frac{1}{\sqrt{2\pi}}\int g{(\omega_{1},\omega^{'}_{1})}e^{-i\omega^{'}_{1}(t-t_{0})}d\omega^{'}_{1} .
\end{equation} 
Note that the action of the input operator causes the reservoir state to collapse to a single photon state, but that the resultant state is still time dependent, due to the presence of $g_{1}(\omega_{1},t)$). The function $g_{1}(\omega_{1},t)$ introduces a memory effect in the reservoir which gives a non-Markovian structure to the final master equations. Finally, we note that for a symmetrized and factorized two-photon spectral envelope we obtain
\begin{equation}
\begin{split}
&\sqrt{\gamma_{iR}}{\rm Tr}_{S\oplus R}\Bigg[[\hat{X}(t),\hat{c}^{\dagger}_{i}(t)]\hat{b}^{(1R)}_{in}(t)\hat{\rho}(t_{0})\Bigg]=\\
&\Omega(t){\rm Tr}_{S}\Bigg[\hat{X}(t_{0})[\hat{c}^{\dagger}_{i},\hat{\rho}_{12}(t)]\Bigg],
\end{split}
\end{equation}
where $\Omega(t)\equiv \sqrt{2\gamma_{iR}}g(t)$. Using this result in the above master equation and replacing $\hat{c}_{i}$ with the atomic lowering operator $\hat{\sigma}^{-}_{i}$, we c obtain the required two-photon bi-directional Fock state master equation for $\hat{\rho}_{s}(t)$. The master equations obeyed by the remaining operators $\hat{\rho}_{ij}$ can also be derived analogously, by using the identity
\be
{\rm Tr}_{S\oplus R}\left[\frac{d\hat{X}(t)}{dt}\hat{\rho}_{ij}(t_{0})\right]={\rm Tr}_{S}\left[\hat{X}(t_{0})\frac{d\hat{\rho}_{ij}(t)}{dt}\right].
\ee

\bibliographystyle{ieeetr}
\bibliography{Article}
\end{document}